\documentclass[num-refs]{wiley-article}
\usepackage{siunitx,amsmath}
\papertype{Research Article}

\title{Estimation of ascertainment bias and its effect on power in clinical trials with time-to-event outcomes}

\author[1]{Erich J.~Greene}
\author[1]{Peter Peduzzi}
\author[2,1]{James Dziura}
\author[1]{Can Meng}
\author[3]{Michael E.~Miller}
\author[4]{Thomas G.~Travison}
\author[1]{Denise Esserman}

\affil[1]{Department of Biostatistics, Yale School of Public Health, New Haven, CT, 06510, USA}
\affil[2]{Department of Emergency Medicine, Yale School of Medicine, New Haven, CT, 06510, USA}
\affil[3]{Department of Biostatistics and Data Science, Wake Forest School of Medicine, Winston-Salem, NC, 27101, USA}
\affil[4]{Marcus Institute for Aging Research, Hebrew SeniorLife, Harvard Medical School, Boston, MA, 02115, USA}

\corraddress{Erich J. Greene, Yale Center for Analytical Sciences, 300 George St Suite 511, New Haven, CT, 06511, USA}
\corremail{erich.greene@yale.edu}

\fundinginfo{Patient-Centered Outcomes Research Initiative (PCORI) and the National Institute on Aging (NIA), Grant Number U01 AG048270; National Center for Advancing Translations Science (NCATS), CTSA Grant Number UL1 TR000142; NIA, Center Core Grant Number P30 AG031679}

\runningauthor{Greene et al.}

\begin{document}

\maketitle

\begin{abstract}
While the gold standard for clinical trials is to blind all parties -- participants, researchers, and evaluators -- to treatment assignment, this is not always a possibility.  When some or all of the above individuals know the treatment assignment, this leaves the study open to the introduction of post-randomization biases.  In the Strategies to Reduce Injuries and Develop Confidence in Elders (STRIDE) trial, we were presented with the potential for the unblinded clinicians administering the treatment, as well as the individuals enrolled in the study, to introduce ascertainment bias into some but not all events comprising the primary outcome. In this manuscript, we present ways to estimate the ascertainment bias for a time-to-event outcome, and discuss its impact on the overall power of a trial versus changing of the outcome definition to a more stringent unbiased definition that restricts attention to measurements less subject to potentially differential assessment. We found that for the majority of situations, it is better to revise the definition to a more stringent definition, as was done in STRIDE, even though fewer events may be observed.

\end{abstract}

\section{Motivation}

\subsection{Introduction}
Although the gold standard for clinical trials is to blind participants, clinical personnel/researchers, and evaluators to treatment assignment, not all designs and treatments allow for blinding of participants and/or researchers. For example, it is not possible to blind participants and clinicians in a trial comparing grouped versus individual treatment for clinical depression. Under these circumstances, trials have the potential to introduce post-randomization detection or ascertainment bias, defined as ``systematic differences between groups in how outcomes are determined.''\cite{higgins2011cochrane}  It is thus important to take appropriate measures to mitigate this type of bias and obtain valid treatment comparisons.

\subsection{Motivating Example} \label{subsection:stride}

The potential for ascertainment bias surfaced in the Strategies to Reduce Injuries and Develop Confidence in Elders (STRIDE) trial,\cite{Bhasin2017,Bhasin2020} funded primarily by the Patient Centered Outcomes Research Institute with additional support from the National Institute on Aging at the National Institutes of Health. Briefly, STRIDE was a cluster-randomized clinical trial designed to test a multicomponent intervention vs. usual care to prevent serious fall injuries. The trial enrolled 5,451 patients from 86 clinical practices in 10 healthcare systems over 20 months (August 1, 2015 to March 31, 2017) and followed them for up to 44 months (ending March 31, 2019).  The primary outcome was time from enrollment to first serious fall injury (SFI), based on self-report data and subsequent verification through the medical record.\cite{Ganz2019}  Multiple papers about the design,\cite{Bhasin2017} recruitment,\cite{Gill2018} intervention,\cite{Reuben2017} outcome adjudication,\cite{Ganz2019} retention,\cite{Gill2020} and primary results\cite{Bhasin2020} of the trial have been published, and interested readers are referred to them for more detailed information beyond the scope of this paper's discussion.

The trial was unblinded, and the intervention was administered by nurse falls care managers (FCM) who interacted with participants enrolled at intervention sites, but not those at sites randomized to standard of care. Because the initial study definition of a serious fall injury included seeking medical attention, there was the potential for interactions with the FCM to create ascertainment bias by sensitizing patients randomized to the intervention to the potential sequelae of falls (potentially making them more likely than control participants to seek medical attention) and for the FCM to refer participants to seek medical attention if they reported a fall. Since this type of referral bias could not occur in the control arm, there was the potential for more falls to be reported in the intervention arm, leading to a dilution of the treatment effect.

\begin{figure}[tbp]
    \centering
    \includegraphics[page=1,width=\linewidth,keepaspectratio]{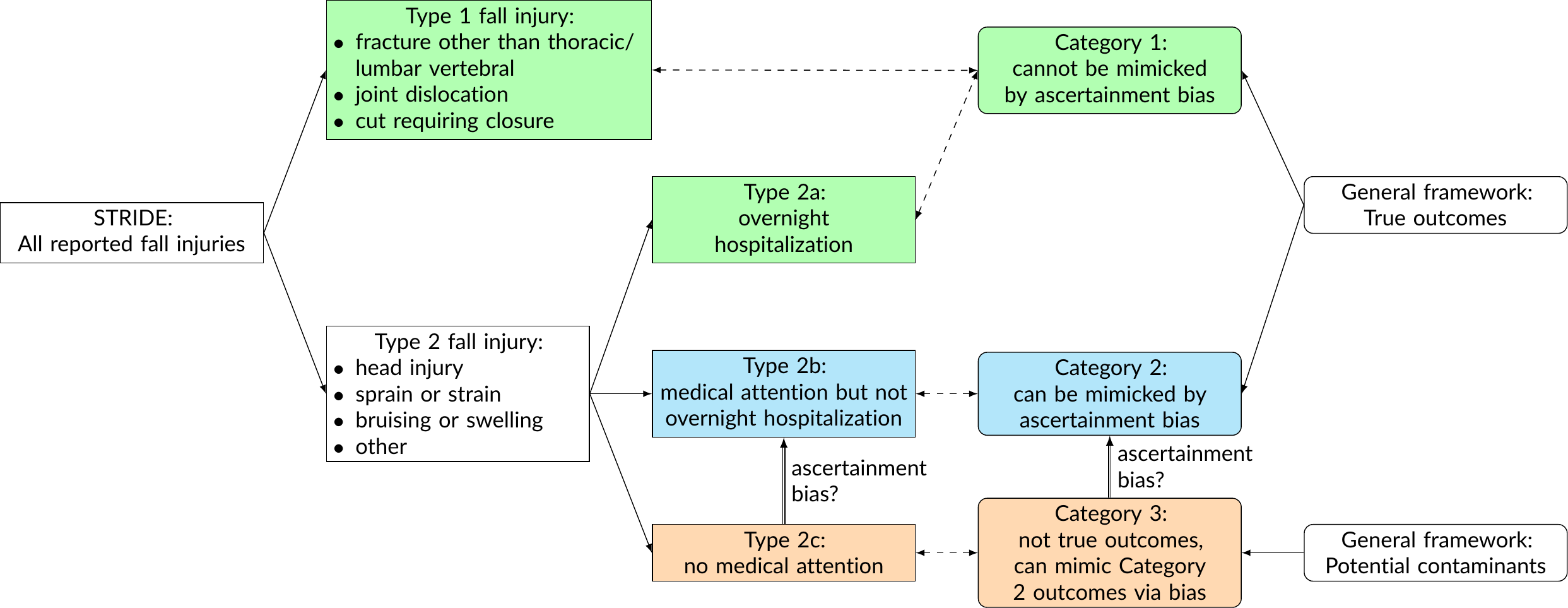}
    \caption{Definitions of and relationships between types of STRIDE fall injuries (Section \ref{subsection:stride}, rectangles flowing from the left) and generalized outcome categories (Section \ref{sect:overview}, ovals flowing from the right).  Green boxes are included in both the original/protocol outcome definition and the restricted/revised outcome definition, blue boxes are included in the original/protocol outcome definition but excluded from the restricted/revised outcome definition, and orange boxes are excluded from both outcome definitions.}
    \label{fig:flowchart}
\end{figure}

The original protocol definition of the primary outcome was, ``a fall injury leading to medical attention, including non-vertebral fractures, joint dislocation, head injury, lacerations, and other major sequelae (e.g., rhabdomyolysis, internal injuries, hypothermia).''  Under this definition, seeking medical attention was intended to affirm the seriousness of the injury.  Fall-related injuries were classified into two types:
\begin{description}
\item[Type 1:] Fracture other than thoracic/lumbar vertebral; joint dislocation; or cut requiring closure; and
\item[Type 2:] Head injury; sprain or strain; bruising or swelling; or other,
\end{description}
with Type 2 fall-related injuries further divided into the following three subtypes:
\begin{description}
\item[Type 2a:] Type 2 fall-related injury resulting in an overnight hospitalization;
\item[Type 2b:] Type 2 fall-related injury resulting in medical attention but not an overnight hospitalization; and
\item[Type 2c:] Type 2 fall-related injury not resulting in medical attention.
\end{description}
(See the left portion of Figure \ref{fig:flowchart} for a visual representation of these definitions.)  The original protocol definition of the primary outcome counted fall-related injuries of Types 1, 2a, and 2b.

We considered the more serious injuries, Type 1 and Type 2a, less susceptible to potential bias because of their comparatively definitive nature and severity. But since interacting with the FCM could lead participants to seek medical attention for less serious injuries, thus converting Type 2c injuries (which are not counted in the original protocol definition) to Type 2b (which are), there was the potential for the FCM (or participants) to introduce bias in the intervention arm ascertainment of events. Note that it was not possible to adjudicate Type 2c events (with no medical attention received, there were no medical records available to adjudicate), though Types 1, 2a and 2b could be adjudicated,\cite{Ganz2019} so it was necessary to use self-reported rather than adjudicated events to estimate the degree of ascertainment bias.

\subsection{Overview} \label{sect:overview}

In this article, we examine the situation in which an intervention is expected to affect the risk of an outcome, but the outcome is operationalized in a way that makes some, but not all, cases in the intervention arm subject to ascertainment bias.  For concreteness, we generally write for the case in which ascertainment bias can produce spurious extra events in the intervention arm, and a useful framework is to consider three categories of possible intervention outcomes:
\begin{description}
\item[Category 1:] true outcome events that cannot be contaminated by ascertainment bias (e.g. Type 1 and Type 2a in STRIDE),
\item[Category 2:] true outcome events that can be mimicked by non-outcome events via bias (e.g. Type 2b in STRIDE), and
\item[Category 3:] events that should not meet the outcome definition but can mimic Category 2 outcomes via bias (e.g. Type 2c in STRIDE).
\end{description}
(See Figure \ref{fig:flowchart} for a visual representation of these categories and their relationships to the STRIDE injury types described in Section \ref{subsection:stride}.)

Throughout the manuscript, for simplicity, we will consider the above categories and scenario; however, it should be noted that in situations where ascertainment bias instead masks true outcome events in the intervention arm, the derivations presented remain applicable,  with the event categories defined slightly differently:
\begin{description}
\item[Category 1:] true outcome events that cannot be contaminated by ascertainment bias,
\item[Category 2:] events that should meet the outcome definition but may fail to in the presence of ascertainment bias, and
\item[Category 3:] events that should not meet the outcome definition and can be mimicked by Category 2 outcomes via bias.
\end{description}

Furthermore,  the mathematical development below continues to apply even if both forms of ascertainment bias are present at once, with Category 2 generally being events that should be included in the outcome but may not be and Category 3 generally being events that should not be included in the outcome but may be.  The key assumption, elaborated below in Section \ref{sect:methods}, is that the ascertainment bias acts by appearing to shuffle events between these two categories.

Given the above event category taxonomy, it is possible to define an alternative outcome definition that's impervious to ascertainment bias by restricting it to include only Category 1 events.  If the hypothesized effect is a risk reduction while the ascertainment bias masks true Category 2 intervention events, or if the hypothesized hazard ratio is greater than 1 while the ascertainment bias introduces spurious Category 2 intervention events, then the bias is away from the null, making such a definition change imperative.  Of more interest is the case where a risk reduction is hypothesized, but the ascertainment bias introduces spurious Category 2 intervention events (or a hypothesized hazard ratio greater than 1 with ascertainment bias masking true Category 2 intervention events).  Here the bias is toward the null, causing a loss in power. The alternative definition also results in a loss of power, only it is through a reduction in the number of observable outcomes by defining a more stringent (i.e., less prone to bias) outcome definition.

While there is an argument to be made for always choosing an outcome definition that allows the trial to produce an unbiased estimate of the treatment effect, given the direction of the bias (toward the null), this decision may not always be in the best interest of the clinical trial and its participants, especially for ongoing clinical trials.  Stakeholders may be forced to weigh the impact and magnitude of the ascertainment bias against the reduction in power from restricting the outcome definition. And the difficulty is magnified in ongoing trials, when the body making the decision about adapting the outcome definition will need to be presented information about the severity of ascertainment bias in a manner that maintains blinding with respect to the treatment effect.  

The methods described in this article were motivated by a need to perform an interim analysis to assess the level of ascertainment bias in STRIDE, estimate its effect on power under the original protocol definition, and estimate the effect on power of revising the outcome definition to include only Type 1 and Type 2a fall-related injuries and to present the results to a blinded ad hoc advisory committee for a decision about whether to adapt the primary outcome. 

The rest of the manuscript is laid out as follows: In Section \ref{sect:methods} we describe the methods used.  We give the overall approach (Section \ref{sect:approach}), define the assumptions (Section \ref{sect:assumptions}), and describe the methods for estimating the effect of ascertainment bias on power (Section \ref{sect:estbias}) and the methods for estimating the effect of revising the outcome on power (Section \ref{sect:estalt}).   We then present the results of the STRIDE example (Section \ref{sect:results}) and perform sensitivity analyses and additional studies of the results (Section \ref{sect:sims}) through simulation.  Finally, we present a discussion, some limitations, and potential future directions in Sections \ref{sect:discussion} and \ref{sect:lims}.

Note that at the time of the decision regarding adaptation of the STRIDE primary outcome, the planned maximum follow-up time was 40 months (it was later extended to the 44 months noted above), so a 40-month trial duration is used in Sections \ref{sect:results} and \ref{sect:sims}.

\section{Methods} \label{sect:methods}

\subsection{Overall approach} \label{sect:approach} 

We consider the scenario in which the outcome of interest is time-to-event, the protocol outcome definition is prone to ascertainment bias, and the bias will  increase the number of observed events in the intervention arm, in turn increasing the observed event rate in the intervention arm.  Since the control arm is unaffected, this will also increase the total number of observed events, resulting in the hazard ratio (intervention relative to control) moving toward 1 (assuming the intervention is effective).

We calculate power using the log-rank test method of Schoenfeld,\cite{Schoenfeld1983}
\begin{equation*} \label{eq:power}
\Phi(z_{1-\beta}) = \Phi\left(\tfrac{1}{2}\sqrt{E} \left| \ln H \right| - z_{1-\alpha/2}\right),
\end{equation*}
where $E$ is the total number of events,  $H$ is the hazard ratio (intervention relative to control), $1-\beta$ is the power, $\alpha$ is the type I error rate, and $\Phi(\cdot)$ is the cumulative normal probability.  From this equation, we see that increasing the number of events ($E$) increases power, but reducing the impact of the intervention (making $H$ closer to 1, or $\left| \ln H \right|$ closer to 0) decreases power. Both effects need to be accounted for; therefore, we:
\begin{enumerate}
\item Estimate the degree of ascertainment bias in Category 2 intervention events (Section \ref{sect:estE}),
\item Estimate how this bias effects the overall number of intervention events (Section \ref{sect:estE}),
\item Use the bias-adjusted number of intervention events to estimate the effective hazard ratio (Section \ref{sect:estH}), and
\item Use the bias-adjusted event total and effective hazard ratio to estimate power (Section \ref{sect:estpower}).
\end{enumerate}

Conversely, if we assume that the true hazard ratio is the same across event types, restricting the outcome definition to Category 1 will have no impact on the hazard ratio. Thus, when evaluating a change in the outcome definition, we only need to account for decreasing the numbers of observed events in both arms (since Category 2 events will no longer be counted).  This is described in Section \ref{sect:estalt}.

\subsection{Assumptions} \label{sect:assumptions}

While this article is written for the context of a clinical trial or other two-arm study with a total duration $T$ and a time-to-first-event outcome in the presence of a semi-competing risk, it is possible to generalize the approach for other designs and scenarios.

If we assume uniform enrollment over the initial fraction $\mu$ of the total study period $T$, an outcome hazard of $\lambda$ in the control arm, and a competing risk hazard of $\gamma$, then for a given hazard ratio $H$ (leading to an outcome hazard of $H\lambda$ in the intervention arm), events in the control ($E_C$) and intervention arms ($E_I$) will be generated (see Appendix \ref{app:numevents} online for a derivation) according to
\begin{align}
E_C &= \frac{N_C^{*}\lambda}{\lambda + \gamma} \left[ 1 - \frac{e^{-(1-\mu)T(\lambda + \gamma)}-e^{-T(\lambda + \gamma)}}{\mu T(\lambda + \gamma)} \right] , \label{eq:Ec} \\
E_I &= \frac{N_I^{*} H\lambda}{H\lambda + \gamma} \left[ 1 - \frac{e^{-(1-\mu)T(H\lambda + \gamma)}-e^{-T(H\lambda + \gamma)}}{\mu T(H\lambda + \gamma)} \right], \label{eq:Ei}
\end{align}
where the control and intervention sample sizes $N_C^{*}$ and $N_I^{*}$ have been adjusted for factors such as expected loss to follow-up, design effect/variance inflation, and interim looks (see Section \ref{sect:results}).

To make the bias problem tractable, we make several assumptions:
\begin{enumerate}
\item The hypothesized hazard ratio (the ratio of intervention to control hazards the investigators expect to observe in the absence of ascertainment bias, i.e., the hypothesized true treatment effect), $H^{hyp}$, is the same for all event categories $e$ (1, 2, and 3, even though Category 3 is not part of the outcome).  It follows (see Appendix \ref{app:justifyconstantx} online for details) that there is a single constant $\kappa$ relating true control and true intervention events such that 
\begin{equation} \label{EQ:CONSTANTEFFECT}
\forall e \in \lbrace 1,2,3 \rbrace: E_{Ie}^{true} = \kappa E_{Ce}^{true} .
\end{equation}

\item Ascertainment bias acts only in the intervention arm and only on Categories 2 and 3, causing some true Category 3 events to become observed intervention Category 2 events and/or vice versa.  This implies
\begin{align*}
E_{I1}^{obs} &= E_{I1}^{true} ,  \\
E_{I2}^{obs} + E_{I3}^{obs} &= E_{I2}^{true} + E_{I3}^{true} , 
\end{align*}
and
\begin{equation} \label{eq:Ec_is_true}
\forall e \in \lbrace 1,2,3 \rbrace: E_{Ce}^{obs} = E_{Ce}^{true} = E_{Ce} .
\end{equation}
(Since we assume the observed and true events are equal in the control arm, no superscript is needed below for the control events.)
\end{enumerate}

\subsection{Estimating the effect of ascertainment bias on power} \label{sect:estbiaseff}

\subsubsection{Estimating the effect on number of events} \label{sect:estE}

We first estimate the degree of ascertainment bias in Category 2 intervention events.
Let $\rho_I$ and $\rho_C$ be the proportion of combined Category 2 and 3 events that are observed in Category 2 in the intervention and control arms, respectively, and let $B$ be the ratio of these proportions, such that:
\begin{align}
\rho_I &= \frac{E_{I2}^{obs}}{E_{I2}^{obs}+E_{I3}^{obs}} \label{eq:rhoi} \\
\rho_C &= \frac{E_{C2}}{E_{C2}+E_{C3}} \label{eq:rhoc} \\
B &= \frac{\rho_I}{\rho_C}. \label{eq:bdef}
\end{align}

Through substitution and some algebra, we find that $B$ can also be defined as the inflation in observed intervention Category 2 events due to ascertainment bias:
\begin{equation*}
B = \frac{E_{I2}^{obs}}{E_{I2}^{true}}.
\end{equation*}
It is important to note that $B$ does not provide information about the actual treatment effect and can therefore be presented to a blinded decision-making panel.

We can therefore express the excess Category 2 events in terms of $B$ and the observed number of events in the intervention arm: 
\begin{align*}\begin{split}
E_{I2}^{excess} &= E_{I2}^{obs} - E_{I2}^{true} \\
&= (B-1) E_{I2}^{true} \\
&= (1-\tfrac{1}{B}) E_{I2}^{obs} .
\end{split}\end{align*}

Next, we estimate how this bias effects the overall number of intervention events.
Let $P$ be the fraction of control events (Categories 1 and 2) that are in Category 2:
\begin{equation}
P = \frac{E_{C2}}{E_C} = \frac{E_{C2}}{E_{C1} + E_{C2}} \label{eq:Pdef}
\end{equation}
By Equation \eqref{EQ:CONSTANTEFFECT}, this will also be the expected ratio among the true intervention events, such that
\begin{align*}
E_{I2}^{true} &= P E_I^{true} \label{eq:2bprop} \\
E_{I1}^{true} &= (1-P) E_I^{true} . 
\end{align*}

Therefore, through substitution and algebra, we obtain $E_I^{obs}$ as
\begin{equation*}
E_I^{obs} = E_I^{true} [1 + P(B-1)] = kE_I^{true}, \label{eq:eventscale}
\end{equation*}
where 
\begin{equation}
k= 1+P(B-1) \label{eq:kdef}
\end{equation}
is the proportionality factor, and the excess number of events in the intervention arm, $E_I^{excess}$, is defined as
\begin{align*}
\begin{split}
E_I^{excess} &= (k-1) E_I^{true} \\
&= (1-\tfrac{1}{k}) E_I^{obs} .
\end{split}\end{align*}

Finally, if the true numbers of events are given by equations \eqref{eq:Ec} and \eqref{eq:Ei}, the number of observed intervention events will be
\begin{equation} \label{eq:obsEi}
E_I^{obs} = kE_I^{true} = \frac{k N_I^* H^{hyp}\lambda}{H^{hyp}\lambda + \gamma} \left[ 1 - \frac{e^{-(1-\mu)T(H^{hyp}\lambda + \gamma)}-e^{-T(H^{hyp}\lambda + \gamma)}}{\mu T(H^{hyp}\lambda + \gamma)} \right] .
\end{equation}

\subsubsection{Estimating the effect on the hazard ratio} \label{sect:estH}

Working from Equation \eqref{eq:Ei}, we can also express the observed number of intervention events $E_I^{obs}$ in terms of an effective hazard ratio (ratio of intervention to control hazards in the presence of ascertainment bias) $H^{eff}$:
\begin{equation} \label{eq:defHeff}
E_I^{obs} = \frac{N_I^* H^{eff}\lambda}{H^{eff}\lambda + \gamma} \left[ 1 - \frac{e^{-(1-\mu)T(H^{eff}\lambda + \gamma)}-e^{-T(H^{eff}\lambda + \gamma)}}{\mu T(H^{eff}\lambda + \gamma)} \right] .
\end{equation}
Combining Equations \eqref{eq:obsEi} and \eqref{eq:defHeff}, we obtain an expression relating the effective hazard ratio $H^{eff}$ to the hypothesized hazard ratio $H^{hyp}$, the control event hazard $\lambda$, the competing risk hazard $\gamma$, the study duration $T$, and the recruitment period fraction $\mu$ (but not sample size or other factors that might affect the number of events independent of the hazard ratio):
\begin{equation} \label{EQ:HMONSTER}
\frac{H^{eff}\lambda}{H^{eff}\lambda+\gamma}\left[ 1 -\frac{e^{-(1-\mu)T(H^{eff}\lambda + \gamma)}-e^{-T(H^{eff}\lambda + \gamma)}}{\mu T(H^{eff}\lambda + \gamma)} \right] = 
\frac{kH^{hyp}\lambda}{H^{hyp}\lambda+\gamma}\left[ 1 - \frac{e^{-(1-\mu)T(H^{hyp}\lambda + \gamma)}-e^{-T(H^{hyp}\lambda + \gamma)}}{\mu T(H^{hyp}\lambda + \gamma)} \right] .
\end{equation}

This equation does not have a closed-form solution for $H^{eff}$, but both Taylor series (expanding in powers of $T(H^{eff}\lambda + \gamma)$) and numerical approximations are described in Appendix \ref{app:solve4H} online.  A first-order approximation is
\begin{equation*}
H^{eff}_{[1]} = \frac{kH^{hyp}}{\left( 1-\tfrac{\mu}{2} \right) T (H^{hyp}\lambda+\gamma)} \left[ 1 - \frac{e^{-(1-\mu)T(H^{hyp}\lambda + \gamma)}-e^{-T(H^{hyp}\lambda + \gamma)}}{\mu T(H^{hyp}\lambda + \gamma)} \right] ,
\end{equation*}
while to second order,
\begin{equation*}
H^{eff}_{[2]} = \frac{\theta}{\lambda} \left( 1 - \sqrt{1 - \phi^2/\theta^2} \right)
\end{equation*}
where
\begin{equation*}
\theta = \frac{1}{2} \left[ \frac{3}{T} \left(\frac{2-\mu}{\mu^2 - 3\mu +3}\right) - \gamma \right]
\end{equation*}
and
\begin{equation*}
\phi = \frac{1}{T} \sqrt{\left( \frac{6}{\mu^2 - 3\mu +3} \right) \left( \frac{kH^{hyp}\lambda}{H^{hyp}\lambda+\gamma} \right) \left[ 1 - \frac{e^{-(1-\mu)T(H^{hyp}\lambda + \gamma)}-e^{-T(H^{hyp}\lambda + \gamma)}}{\mu T(H^{hyp}\lambda + \gamma)} \right] }.
\end{equation*}
Since the expansion parameter may not be small compared to 1 (e.g., in the STRIDE trial, $T(H^{eff}\lambda + \gamma) > T(H^{hyp}\lambda + \gamma) \approx 0.52$), these approximations are best used as starting points for numerical estimation.

Note that when estimating these quantities from observed data, we can compute the cause-specific hazards ($\lambda$ and $\gamma$) from the observed cause-specific rates (see Appendix \ref{app:realrates} online for details).

\subsubsection{Estimating the overall effect on power} \label{sect:estpower}

Once $H^{eff}$ has been estimated, the projected power under the protocol definition can be obtained via:
\begin{equation}
\Phi(z_{1-\beta^{prot}}) = \Phi\left(\tfrac{1}{2}\sqrt{E_C + E_I^{obs}} \left| \ln H^{eff} \right| - z_{1-\alpha/2}\right) ,  \label{eq:biaspower}
\end{equation}
with $E_C$ defined as in Equation \eqref{eq:Ec} and $E_I^{obs}$ defined as in Equation \eqref{eq:obsEi}.  (In this equation and in what follows, the superscript $prot$ denotes quantities specific to the protocol definition.)

\subsection{Estimating the effect of redefining the outcome on power} \label{sect:estalt}

If we change our outcome definition to include only Category 1 events, then if the true number of total control events (in both Categories 1 and 2), $E_C$, is given by Equation \eqref{eq:Ec}, the number of control events under the revised definition will be
\begin{equation*}
E_C^{redef} = (1-P)E_C = \frac{(1-P)N_C^*\lambda}{\lambda + \gamma} \left[ 1 - \frac{e^{-(1-\mu)T(\lambda + \gamma)}-e^{-T(\lambda + \gamma)}}{\mu T(\lambda + \gamma)} \right], \label{eq:redefEc}
\end{equation*}
where $P$ is the proportion of control events that are Category 2 (as defined in Equation \ref{eq:Pdef}).  (In this equation and in what follows, the superscript $redef$ denotes quantities specific to the revised definition.)

Since Category 1 events are not subject to ascertainment bias, there is no longer an effect on the hazard ratio, and the number of intervention events under the revised definition will be
\begin{equation*}
E_I^{redef} = \frac{(1-P)N_I^* H^{hyp}\lambda}{H^{hyp}\lambda + \gamma} \left[ 1 - \frac{e^{-(1-\mu)T(H^{hyp}\lambda + \gamma)}-e^{-T(H^{hyp}\lambda + \gamma)}}{\mu T(H^{hyp}\lambda + \gamma)} \right]  = (1-P)E_I , \label{eq:redefEi}
\end{equation*}
with $E_I$ taken from Equation \eqref{eq:Ei}.

Using these redefined expected events,  the projected power can be obtained from the following:
\begin{equation}
\Phi(z_{1-\beta^{redef}}) = \Phi\left(\tfrac{1}{2}\sqrt{E_C^{redef} + E_I^{redef}} \left| \ln H^{hyp} \right| - z_{1-\alpha/2} \right) = \Phi\left(\tfrac{1}{2}\sqrt{[1-P][E_C + E_I]} \left| \ln H^{hyp} \right| - z_{1-\alpha/2}\right) .  \label{eq:redefpower}
\end{equation}

\section{Results: STRIDE Example} \label{sect:results}

This method and the preliminary data obtained from the STRIDE study informed the decision to change the definition of the primary outcome\cite{CTG} to avoid impacts of potential ascertainment bias.  Following from Section \ref{sect:methods}, we demonstrate for the STRIDE trial how to estimate the bias (Section \ref{sect:estbias}; calculation of confidence intervals is discussed in Appendix \ref{app:calcCI} online), adjust the sample sizes to obtain $N^*_C$ and $N_I^*$ (Section \ref{sect:effss}), estimate the hazards (Section \ref{sect:hazards}), and estimate the projected number of self-reported events (Section \ref{sect:projsr}).  Since STRIDE adjudicated self-reported outcomes, we present methods to estimate the projected number of adjudicated events in Section \ref{sect:adjudication}. Then, in Section \ref{sect:biashr} we calculate the effective hazard.  All of these pieces are necessary to determine the power of the trial under the protocol and revised outcome definitions in Section \ref{sect:projpower}.

\subsection{Estimation of ascertainment bias} \label{sect:estbias}

Using all available information to determine the degree of observed bias and estimate its effect on study power, we follow the logic of Section \ref{sect:estbiaseff}.  In a snapshot of all self-report data taken on February 22, 2018, the control group  reported a total of 253 Type 2b (Category 2) events and 613 Type 2c (Category 3) events, while the intervention group  reported 263 and 526, respectively, total events of these types.  Using Equations \eqref{eq:rhoi}, \eqref{eq:rhoc}, and \eqref{eq:bdef}, our proportions $\rho$ are thus $\rho_C = \frac{253}{253+613} = 0.292$ and $\rho_I = \frac{263}{263+526} = 0.333$, and our estimate of ascertainment bias is $B = \rho_I / \rho_C = 1.141$ (95\% CI: 0.978--1.304).  Therefore, the inflation in observed Category 2 intervention events due to ascertainment bias in the STRIDE study at the time of data lock was estimated to be 14.1\%.

For first events (the primary outcome of interest), the control group reported 215 Type 1, 55 Type 2a (for a total of 270 in Category 1) and 206 Type 2b (Category 2) first events. Using Equations \eqref{eq:Pdef} and \eqref{eq:kdef}, we can estimate the Category 2 fraction of first events, $P = \frac{206}{270+206} = 0.433$ (95\% CI: 0.388--0.477; recall from Section \ref{sect:estE} that $P$ is defined in terms of observed control events, which are assumed free from ascertainment bias), and the estimated biased-induced inflation, $k = 1 + (0.433)(0.141) = 1.061$ (95\% CI: 0.990--1.132). Hence, we estimate that we are observing 6.1\% more primary outcome events in the intervention arm due to ascertainment bias.

\subsection{Effective sample sizes} \label{sect:effss}

As described in Section \ref{sect:assumptions}, Equations \eqref{eq:Ec} and \eqref{eq:Ei} require effective control and intervention sample sizes that account for loss to follow-up and the variance inflation/design effect.
 
We can project the overall proportion lost in a $T$-month trial to be
$W = 1 - (1 - w)^{T/(12\text{ mo})}$ ,
where $w$ is the observed loss rate (consent withdrawals per person-year of follow-up).  For STRIDE, with $T = 40\text{ mo}$ and an observed loss rate of $w = 0.022$ (at data lock; this rate was similar across treatment arms), we projected the overall proportion lost to be $W = 0.071$.

We estimated the variance inflation/cluster design effects by calculating the ratios of the variances of the estimates of the intervention effect parameters from lognormal frailty and unclustered Cox models (treating competing events as censored) using self-reported events under the protocol and revised definitions.  At data lock, the variance inflation factors were $V^{prot} = 1.0000$ (95\% CI: 1.0000--1.0000) under the protocol definition and $V^{redef} = 1.0475$ (95\% CI: 1.0000--1.1775) under the revised definition.

Finally, effective sample sizes were calculated as $N^* = N(1-W)/V$.  With actual enrollments of $N_C = 2649$ and $N_I = 2802$, our effective sample sizes were $N_C^{*prot} = 2459.6$ and $N_I^{*prot} = 2601.6$ for the protocol definition and $N_C^{*redef} = 2348.0$ and $N_I^{*redef} = 2483.6$ for the revised definition. These adjustments are made because it is important to recognize that in a trial with clustering and loss to follow-up, the effective sample size $N^*$ will be smaller than the observed baseline sample size.

\subsection{Hazards} \label{sect:hazards}

In the STRIDE snapshot, we observed $r_C^{prot} = 0.148$ patient-reported control events per person year of follow-up (PYF) under the protocol definition, $r_C^{redef} = 0.089$ patient-reported control events per PYF under the revised definition, and $d = 0.025$ deaths per PYF.  Using Equations \eqref{eq:h1} and \eqref{eq:h2} from Appendix \ref{app:realrates} online, we calculated our observed hazards to be $\lambda_C^{prot} = 0.0135 \text{ mo}^{-1}$ and $\gamma^{prot} = 0.0023 \text{ mo}^{-1}$ for the protocol definition and $\lambda_C^{redef} = 0.0079 \text{ mo}^{-1}$ and $\gamma^{redef} = 0.0022 \text{ mo}^{-1}$ for the revised definition.  Since the revised definition includes only a subset of events from the protocol definition, the event hazard under the revised definition is smaller, as one would expect.

\subsection{Projected self-reported events} \label{sect:projsr}

Using the hypothesized hazard ratio $H^{hyp} = 0.8$, the trial duration $T = 40\text{ mo}$, and the recruitment fraction $\mu = 0.5$ from the STRIDE design, along with the estimated bias inflation $k$ from Section \ref{sect:estbias},  effective sample sizes from Section \ref{sect:effss}, and hazards from Section \ref{sect:hazards}, we can project the numbers of self-reported control and intervention events under the protocol and revised definitions.
For the protocol definition, we set $H = H^{hyp} = 0.8$, $\lambda = \lambda_C^{prot} = 0.0135\text{ mo}^{-1}$, and $\gamma = \gamma^{prot} = 0.0023\text{ mo}^{-1}$ and use Equation \eqref{eq:Ec} to project $E_C^{prot,sr} = 789.0$ self-reported control events, Equation \eqref{eq:Ei} to project $E_I^{prot,sr,true} = 694.0$ true self-reported intervention events, and Equation \eqref{eq:obsEi} to project $E_I^{prot,sr,obs} = 736.3$ observed self-reported intervention events.
For the revised definition, we set $H = H^{hyp} = 0.8$, $\lambda = \lambda_C^{redef} = 0.0079\text{ mo}^{-1}$, and $\gamma = \gamma^{redef} = 0.0022\text{ mo}^{-1}$ and use Equation \eqref{eq:Ec} to project $E_C^{redef,sr} = 476.1$ self-reported control events and Equation \eqref{eq:Ei} to project $E_I^{redef,sr} = 412.3$ self-reported intervention events.

\subsection{Projected adjudicated outcome events} \label{sect:adjudication}

The primary analysis for STRIDE uses adjudicated rather than self-reported outcomes, so a calculation of projected power needs to account for some fraction of self-reported events not becoming confirmed outcome events.  Though the data snapshot was taken early in the adjudication process, we used the available information from completed cases to estimate the overall confirmation fractions under the protocol and revised definitions.

Notably, the fraction of self-reported events confirmed in adjudication (denoted $A_t$ where $t$ is the event type) varied across event types: $A_1 = 0.966$ of Type 1 self-reported events were confirmed, while only $A_{2a} = 0.667$ of Type 2a and $A_{2b} = 0.771$ of Type 2b were confirmed.  (These fractions were similar across treatment groups.)  Since the revised definition does not include Type 2b events, and 29 of the 206 control-group participants whose first self-reported events were of Type 2b also reported later events of Type 1 (21) or Type 2a (8) that would be first self-reported events under the revised definition, the proportions of self-reported events of each type differed by definition: $P_1^{prot} = \frac{215}{476} = 0.452$, $P_{2a}^{prot} = \frac{55}{476} = 0.116$, and $P_{2b}^{prot} = \frac{206}{476} = 0.433$ under the protocol definition, while $P_1^{redef} = \frac{236}{299} = 0.789$ and $P_{2a}^{redef} = \frac{63}{299} = 0.211$ under the revised.

The overall probability $A$ that a self-reported event will be adjudicated a true outcome can be estimated as
\begin{equation*}
A = Pr(\textrm{confirm}) = \sum_{t \in \textrm{types}} Pr(t) Pr(\textrm{confirm}|t) = \sum_{t \in \textrm{types}} P_t A_t ,
\end{equation*}
so the confirmation probabilities for the two definitions are $A^{prot} = P_1^{prot}A_1 + P_{2a}^{prot}A_{2a} + P_{2b}^{prot}A_{2b} = 0.847$ and $A^{redef} = P_1^{redef}A_1 + P_{2a}^{redef}A_{2a} = 0.903$; note that while the revised definition includes fewer events, those events are more likely to be confirmed in adjudication. We use these confirmation probabilities to convert the projected self-reported events into projected adjudicated outcomes: $E^{adj} = AE^{sr}$.  For the protocol definition, $E_C^{prot,adj} = 668.5$, $E_I^{prot,true,adj} = 588.0$, and $E_I^{prot,obs,adj} = 623.9$ (i.e., 35.9 [6.1\%] extra observed intervention events due to bias-induced inflation).  For the revised definition, $E_C^{redef,adj} = 430.0$ and $E_I^{redef,adj} = 372.4$.

\subsection{Effective hazard ratio under the protocol definition} \label{sect:biashr}

With $H^{hyp} = 0.8$, $T = 40\text{ mo}$, $\mu = 0.5$, $k = 1.061$, $\lambda = \lambda_C^{prot} = 0.0135\text{ mo}^{-1}$, and $\gamma = \gamma^{prot} = 0.0023\text{ mo}^{-1}$, we solve Equation \eqref{EQ:HMONSTER} numerically using the methods of Appendix \ref{app:solve4H} in the online supplement and project a bias-induced dampening of the treatment effect to $H^{eff} = 0.858$.

\subsection{Projected power} \label{sect:projpower}

Using all of the information from Sections \ref{sect:estbias}-\ref{sect:biashr}, we can calculate the power of the trial under the protocol and revised definitions.  For the protocol definition, Equation \eqref{eq:biaspower} tells us that $\Phi(z_{1-\beta^{prot}}) = \Phi\left(\tfrac{1}{2}\sqrt{668.5 + 623.9} \left| \ln 0.858 \right| - z_{.975}\right) = \Phi(0.782)$, corresponding to 78.3\% projected power.  This is lower than the 88.4\% power projected under the revised definition, following from Equation \eqref{eq:redefpower} and $\Phi(z_{1-\beta^{redef}}) = \Phi\left(\tfrac{1}{2}\sqrt{430.0 + 372.4} \left| \ln 0.8 \right| - z_{.975}\right) = \Phi(1.200)$.  Therefore, the recommendation based on these results was to revise the primary outcome definition to the more stringent (less bias prone) definition, as even with the loss of events, it would still result in greater trial power.

\section{Simulations} \label{sect:sims}

These simulations were run and figures produced using \textsf{SAS/STAT} software, Version 14.3 for Windows.\cite{sas_stat_14_3}  Code is provided as an online supplement.

\subsection{STRIDE sensitivity analyses} \label{sect:sens}

In preparing the report for the ad hoc advisory committee, our calculations were based on a hypothesized hazard ratio and on estimates of ascertainment bias, variance inflation, and adjudication confirmation fraction from a snapshot of data in hand.  However, the actual hazard ratio could be different from the hypothesized value, and the snapshot estimates could be unstable.  To address these concerns, we performed sensitivity analyses to examine the effect on power, under both definitions, of variations in these hypothesized and estimated quantities.

Four parameters of interest, the hypothesized hazard ratio ($H^{hyp}$), ascertainment bias ($B$), variance inflation ($V$), and overall adjudication confirmation fraction ($A$), were varied in sensitivity analyses. For each parameter, the power calculations of Section \ref{sect:results} were repeated with that parameter varying across a range of values, holding all other quantities constant and using the values defined in Section \ref{sect:results}. $H^{hyp}$ was varied from 0.70 to 0.90 in increments of 0.002, $B$ was varied from 1.00 to 1.25 in increments of 0.01, $V$ (used as both $V^{prot}$ and $V^{redef}$) was varied from 1.0 to 1.5 in increments of 0.01, and $A$ (used as both $A^{prot}$ and $A^{redef}$) was varied from 0.50 to 1.00 in increments of 0.01.

\begin{figure}[btp]
\centering
\includegraphics{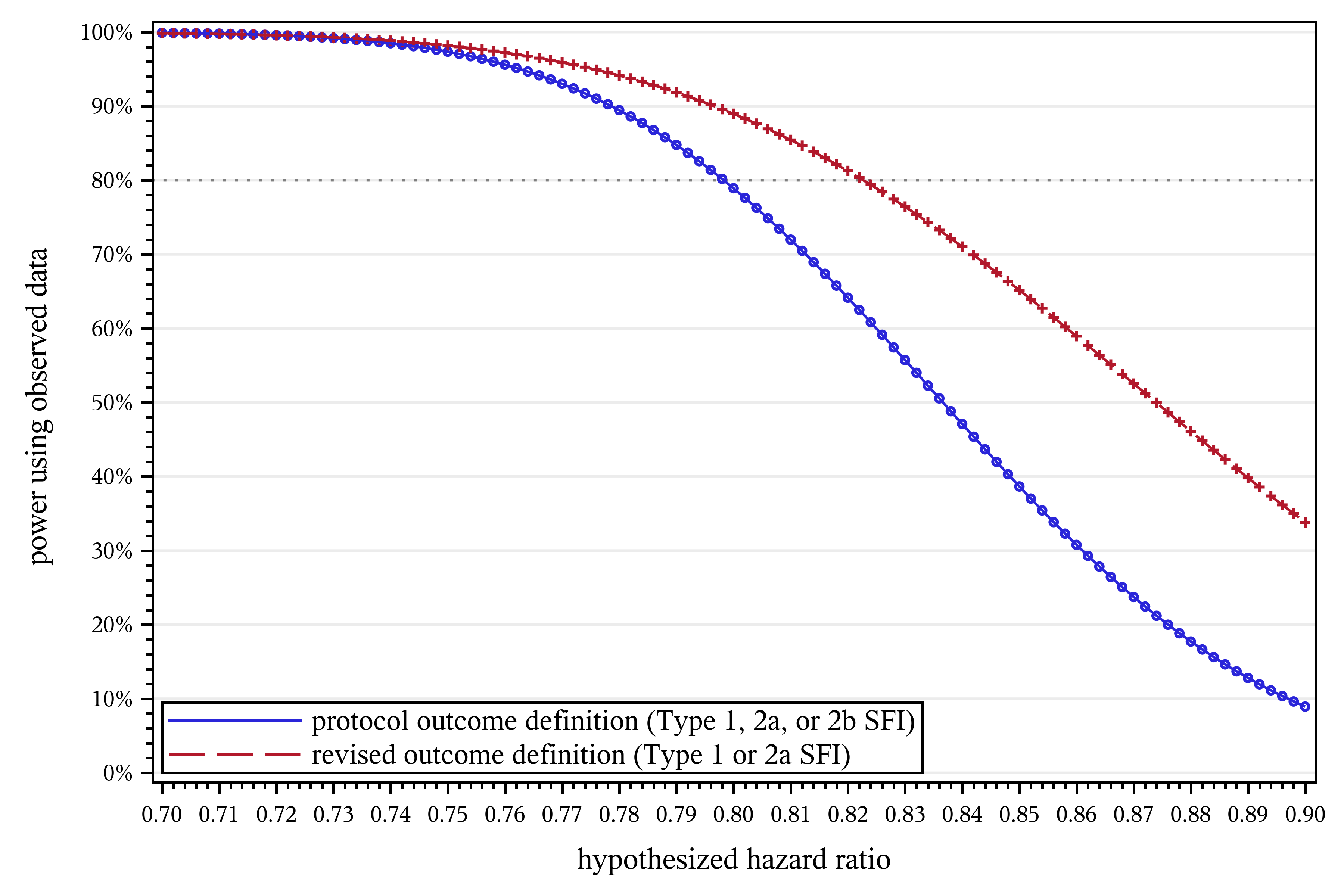}
\caption{Effect of the hypothesized hazard ratio ($H^{hyp}$) on projected power in STRIDE. $N_C = 2649$, $N_I = 2802$, $T = 40 \text{ mo}$, $\mu = 0.5$, $w = 0.022$, $V^{prot} = 1.0000$, $V^{redef} = 1.0475$, $r_C^{prot} = 0.148$, $r_C^{redef} = 0.089$, $d = 0.025$, $A_1 = 0.966$, $A_{2a} = 0.667$, $A_{2b} = 0.771$, $P_1^{prot} = 0.452$, $P_{2a}^{prot} = 0.116$, $P_{2b}^{prot} = 0.433$, $P_1^{redef} = 0.789$, $P_{2a}^{redef} = 0.211$, $B = 1.141$.}
\label{fig:varyH}
\end{figure}

Figure \ref{fig:varyH} demonstrates the effect of the hypothesized hazard ratio on projected power under both the protocol and revised outcome definitions.  Unsurprisingly, power drops for both definitions as the intervention effect weakens (hazard ratio approaches 1).  If the intervention effect is much stronger than anticipated (hazard ratio below about .75 rather than the hypothesized .8), the choice of outcome definition makes no appreciable difference, but the revised definition otherwise retains more power, with the difference becoming more pronounced as the intervention effect weakens.

\begin{figure}[btp]
\centering
\includegraphics{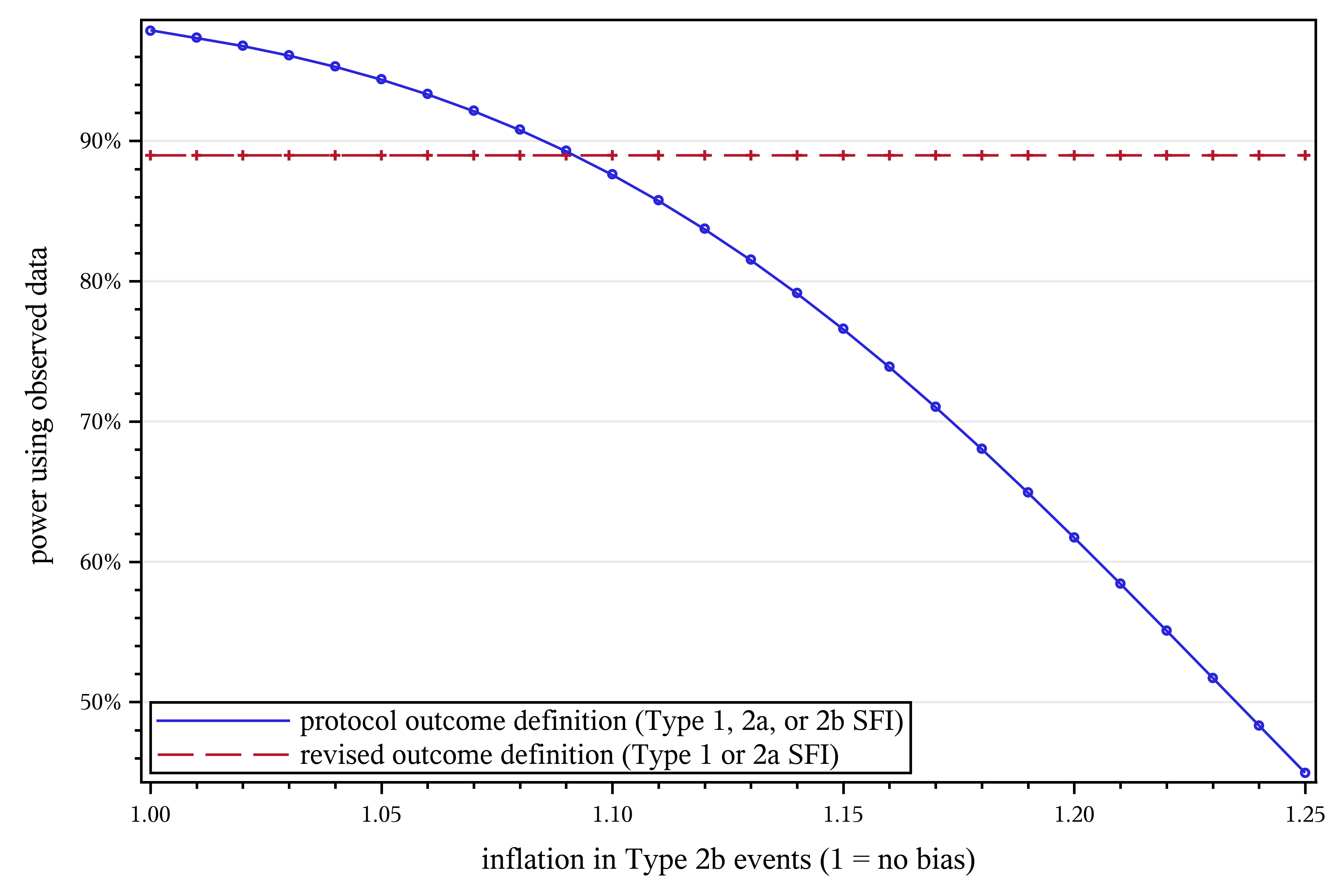}
\caption{Effect of ascertainment bias ($B$) on projected power in STRIDE. $N_C = 2649$, $N_I = 2802$, $T = 40 \text{ mo}$, $\mu = 0.5$, $H^{hyp} = 0.8$, $w = 0.022$, $V^{prot} = 1.0000$, $V^{redef} = 1.0475$, $r_C^{prot} = 0.148$, $r_C^{redef} = 0.089$, $d = 0.025$, $A_1 = 0.966$, $A_{2a} = 0.667$, $A_{2b} = 0.771$, $P_1^{prot} = 0.452$, $P_{2a}^{prot} = 0.116$, $P_{2b}^{prot} = 0.433$, $P_1^{redef} = 0.789$, $P_{2a}^{redef} = 0.211$.}
\label{fig:varyB}
\end{figure}

Figure \ref{fig:varyB} shows how the size of the bias estimate ($B$, the degree of inflation for intervention Type 2b events) affects projected power under both the protocol and revised outcome definitions.  For STRIDE, the power loss from dilution of the treatment effect under the protocol definition matches the loss from adopting a narrower outcome definition at $B = 1.09$; i.e., introducing a 9\% ascertainment bias has the same impact on power as reducing the total number of observed events by 43\% (the fraction of protocol outcome events not counted under the revised definition).

\begin{figure}[btp]
\centering
\includegraphics{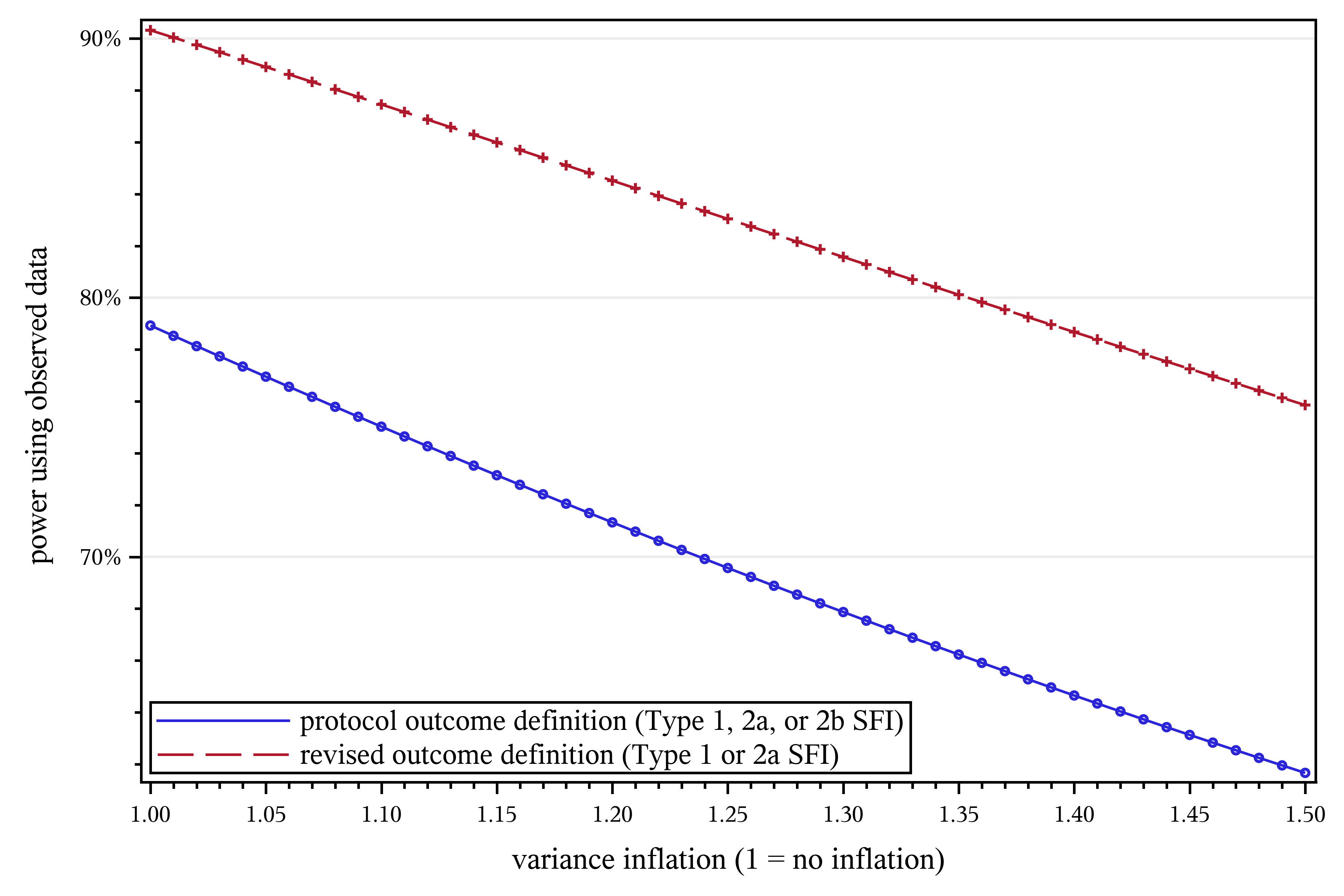}
\caption{Effect of variance inflation ($V$) on projected power in STRIDE. $N_C = 2649$, $N_I = 2802$, $T = 40 \text{ mo}$, $\mu = 0.5$, $H^{hyp} = 0.8$, $w = 0.022$, $r_C^{prot} = 0.148$, $r_C^{redef} = 0.089$, $d = 0.025$, $A_1 = 0.966$, $A_{2a} = 0.667$, $A_{2b} = 0.771$, $P_1^{prot} = 0.452$, $P_{2a}^{prot} = 0.116$, $P_{2b}^{prot} = 0.433$, $P_1^{redef} = 0.789$, $P_{2a}^{redef} = 0.211$, $B = 1.141$.}
\label{fig:varyV}
\end{figure}

\begin{figure}[btp]
\centering
\includegraphics{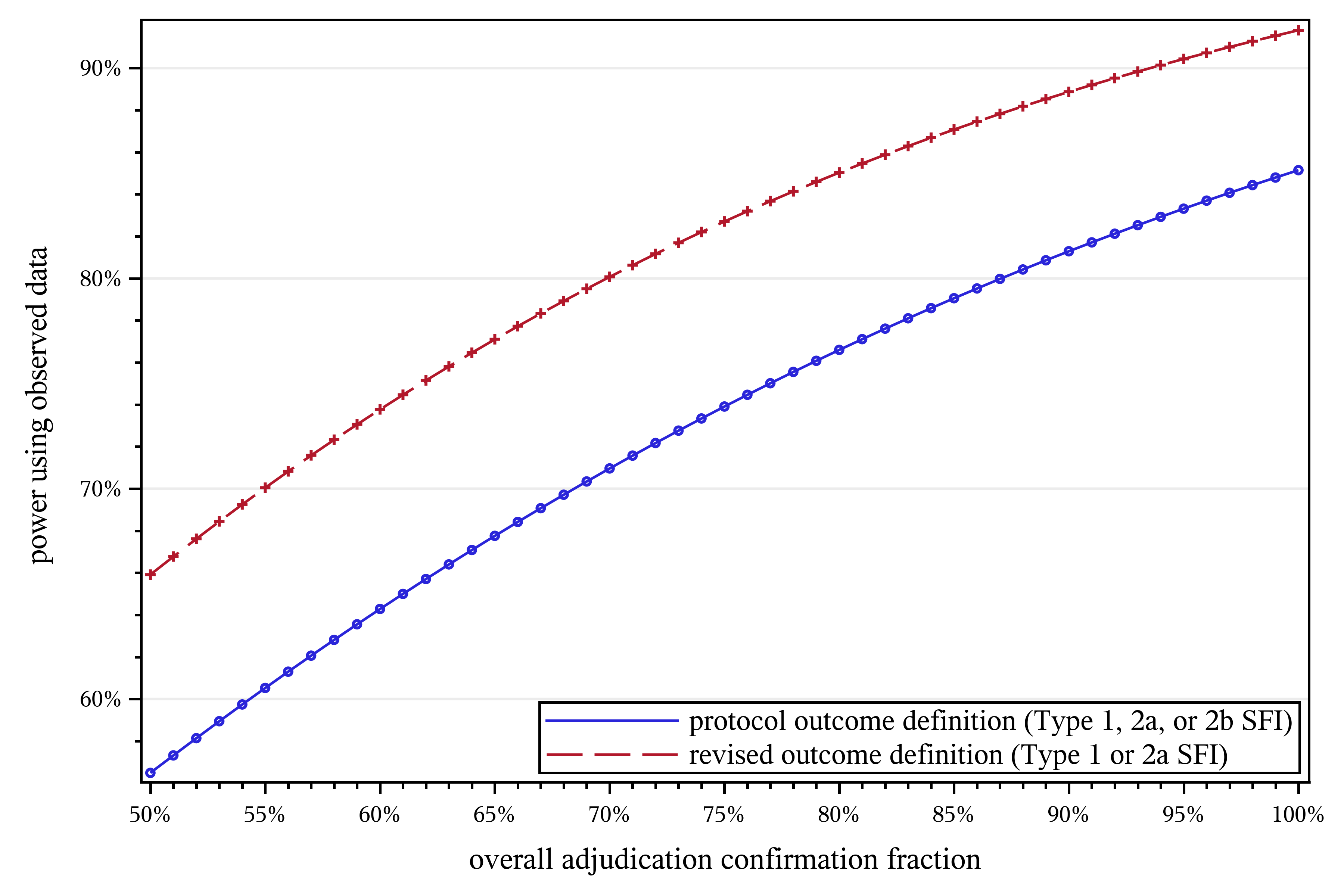}
\caption{Effect of the adjudication confirmation fraction ($A$) on projected power in STRIDE. $N_C = 2649$, $N_I = 2802$, $T = 40 \text{ mo}$, $\mu = 0.5$, $H^{hyp} = 0.8$, $w = 0.022$, $V^{prot} = 1.0000$, $V^{redef} = 1.0475$, $r_C^{prot} = 0.148$, $r_C^{redef} = 0.089$, $d = 0.025$, $P_1^{prot} = 0.452$, $P_{2a}^{prot} = 0.116$, $P_{2b}^{prot} = 0.433$, $P_1^{redef} = 0.789$, $P_{2a}^{redef} = 0.211$, $B = 1.141$.}
\label{fig:varyA}
\end{figure}

Since there are no satisfactory established methods for estimating the design effect for a cluster-randomized survival trial, we employed a model-based estimate of variance inflation in our projections.  This is especially important given the snapshot estimates we obtained could be unstable.  We were particularly interested in the impact of different design effects.  These results are presented in Figure \ref{fig:varyV}; variance inflation naturally has a negative impact on power, but with the revised definition performing generally better than the protocol definition, we gain some insurance against the final design effect being larger than estimated.

Finally, Figure \ref{fig:varyA} shows the effect of the overall adjudication confirmation fraction.  As one would expect, larger confirmation rates result in greater power.  For a given confirmation fraction, the revised definition has greater projected power, and given that Type 1 events were substantially more likely than Type 2b events to be confirmed (Section \ref{sect:adjudication}), the revised definition's higher overall confirmation rate provides a boost on that axis as well.

\subsection{Additional simulations} \label{sect:addsims}

We performed several additional simulations to explore how ascertainment bias might impact the effective hazard ratio and study power across ranges of other parameters.  For simplicity, we return to the full-information event space of Sections \ref{sect:estbias} and \ref{sect:estalt}, without the additional complications of variance inflation and event adjudication that were particular to STRIDE, and calculations parallel those in Section \ref{sect:estpower}.  Parameters held constant in a given simulation were set to STRIDE-like values: unadjusted sample sizes $N_C = N_I = 1611$ (STRIDE's unadjusted targets) in each arm, $T = 40$ month trial, 20 month enrollment period ($\mu = 0.5$), hypothesized hazard ratio $H^{hyp} = 0.8$, observed control-arm first event incidence $r = 14.8\%$ per year (counting both Category 1 and Category 2), observed death incidence $d = 2.5\%$ per year, and $P = 43.2\%$ of full-definition outcomes in Category 2.

In each section that follows, we vary, independently, the ascertainment bias $B$, along with one additional parameter.   In Section \ref{sect:propbias}, we  vary the proportion of Category 2 events ($P$); and  in Section \ref{sect:evrate}, we vary first the control event rate ($r$) and then the follow-up time.  The supplementary figures referenced appear in an online supplement.

\subsubsection{Impact of varying the proportion of events prone to bias} \label{sect:propbias}

The total impact of ascertainment bias arises from two sources: The proportion $P$ of true events that can be mimicked through bias (i.e., that are in Category 2) and the factor $B$ by which the observed number of such events is inflated.  In this simulation, we allowed these two parameters to vary independently; $P$ varied from 0 to 1 and $B$ varied from 0.85 to 1.25, both in increments of .05.

Supplementary Figure \ref{fig:HvaryPB} shows how these variations impact the effective hazard ratio ($H^{eff}$).  $B = 1$ indicates no ascertainment bias and thus no effect on the hazard ratio.  As $B$ drops below 1, ascertainment bias reduces rather than increases the number of observed intervention events, moving the hazard ratio further below 1 and biasing study results against the null and towards the alternative.  For $B > 1$, the dilution of the treatment effect becomes more severe as $B$ and $P$ increase, and in the upper right area of the figure, we see that a combination of more bias-induced event inflation and more  events prone to bias can push the hazard ratio above 1, reversing the apparent direction of the intervention effect.

Supplementary Figure \ref{fig:PvaryPB} demonstrates the impact varying $P$ and $B$ have on study power; again, $B = 1$ indicates no ascertainment bias and no effect on power.  As $B$ drops below 1, study power does increase, but since ascertainment bias is away from the null, this does not reflect a true improvement in sensitivity.  For $B > 1$, study power decreases as either aspect of bias (inflation or prevalence) worsens; greater prevalence also leads to a greater reduction in power under the revised (i.e. more restrictive) definition, as there are fewer unbiased (Category 1) events to retain. Of note, the rebound in power at the high-$P$ end of the $B = 1.25$ curve comes when the effective hazard ratio crosses 1 (see Figure \ref{fig:HvaryPB}); beyond that point, the observed intervention effect becomes opposite to its true direction, with even stronger bias pushing the trial result further away from the null in the wrong direction.

\subsubsection{Impact of varying the event rate and the trial duration} \label{sect:evrate}

Supplementary Figures \ref{fig:HvaryRB} and \ref{fig:PvaryRB} demonstrate how changes in the outcome event rate impact the effective hazard ratio and projected power, respectively, across variations in bias. The control event rate was varied from 0.025 to 0.7 in increments of 0.025, while $B$ was varied from 0.85 to 1.25 in increments of 0.05.  In the presence of ascertainment bias, power  initially increases with the event rate, but as the event rate grows, the slope inverts and power is lost as the inflation in observed intervention events overwhelms the benefit of having more events overall.  Eventually, as the inflation in observed intervention events pushes the effective hazard ratio past 1, power again increases with event rate, but toward a conclusion opposite to the true intervention effect.

Supplementary Figures \ref{fig:HvaryTB} and \ref{fig:PvaryTB} show how changes in the overall trial duration impact the effective hazard ratio and projected power, respectively, across varying bias.  The enrollment period was held fixed at 20 months (for any given simulation, $\mu = \tfrac{20\text{ mo}}{T}$ with $T$ varying from 20 to 160 months in 5 month increments), so these simulations represent the effect of varying follow-up time.  The effect of extending follow-up is similar to the impact of a higher control event rate:  generating additional events is initially helpful, but in the presence of ascertainment bias, the inflation of intervention events eventually becomes the dominant factor.  Perhaps counterintuitively, it is possible for extending follow-up time to exacerbate rather than ameliorate the loss of power due to ascertainment bias.

\section{Discussion} \label{sect:discussion}

Our approach is akin to an interim analysis that uses updated event rates (but not updated estimates of the treatment effect) to look at power, but with the addition of estimating and accounting for the degree of ascertainment bias in the outcome. Though the calculations should be performed by an unblinded statistician, the results can be shared with and evaluated by blinded investigators and decision-makers.

Any bias can impact a study and lead to erroneous results; therefore, it is necessary to mitigate biases. Specifically, a study with ascertainment bias prevents investigators from obtaining a true estimate of the treatment effect. As was shown with the STRIDE example and through the simulation studies, even modest amounts of bias can result in loss of power.  Even more important is that for most scenarios, the power reduction from introducing bias is worse than the power reduction from creating a stricter outcome definition (i.e., removing events prone to bias from the outcome definition results in greater efficiency while reducing error).  In the STRIDE example, the results would most likely be an under-powered study.  While this would be an unfortunate event, we saw that if the bias is severe enough, we could end up with an even more detrimental result -- concluding a result in the completely opposite direction.

We again note that considering the relative efficiency of broader but biased versus narrower but unbiased outcome definitions, as we have here, is only appropriate in situations where any potential ascertainment bias is toward the null.  If there is ascertainment bias toward the alternative, restricting the outcome definition is necessary to maintain control over type I error.

\section{Limitations and Future Work} \label{sect:lims}

The current method requires us to assume that the intervention being investigated affects all outcomes equally, including Category 3 events that aren't included in any version of the outcome definition.  It is possible for an intervention to operate selectively on different types of events, becoming more or less effective for more or less severe events, and the issue may be exacerbated if the intervention doesn't target Category 3 events.

The current method also requires us to assume that ascertainment bias is limited to the intervention arm, making it possible to use the control arm to estimate the relative proportions of event types in the absence of bias.  While this assumption is true for our motivating example, which uses standard of care in the control arm and has no mechanism for introducing this type of bias, there could be trials where ascertainment bias operates in both arms.  In such cases, if the other assumptions are met, the bias parameter $B$ can be interpreted as the ratio of the effect of bias in the intervention arm to its the effect in the control arm, but with Equation \eqref{eq:Ec_is_true} no longer holding, the relationship of $B$ to the numbers of observed events and the power of the trial would no longer be evident.

For simplicity, we have assumed uniform enrollment, which is often done in pre-trial power calculations but may not be true in practice.  To account for a different enrollment scheme, it is possible to return to Equation \eqref{eq:nonuniform} in Appendix \ref{app:numevents} online and use a different parametric distribution $g(\tau)$ of enrollment times.  If the resulting integral has a closed-form solution, it will lead to analogs of Equations \eqref{eq:Ec} and \eqref{eq:Ei} for the numbers of events in each arm, and from there, one can follow the development of Section \ref{sect:estH} to reach a new version of Equation \eqref{EQ:HMONSTER} relating $H^{eff}$ to $H^{hyp}$ and $k$.

STRIDE was a cluster-randomized trial utilizing covariate-constrained randomization\cite{Greene2017,Moulton2004} to achieve balance between the treatment arms, and the primary outcome was analyzed as a marginal effect across the entire study. Thus, we did not consider imbalances between the intervention and control arms in the determination of ascertainment bias.  However, situations may arise where it is desirable to account for covariate imbalance between treatment groups or heterogeneity at the center, cluster, or individual level that could lead to differential susceptibility to ascertainment bias.  It is possible to stratify on one or more variables of concern and carry through the calculations in this paper to produce bias estimates for each stratum, or even to calculate expected observed events and effective hazard ratios for each stratum.  However, these single-stratum estimates will be less stable (and more sensitive to outliers) than the direct unstratified estimate, and methods for combining the single-stratum results into overall projections of study power are a topic for further research.

While obtaining estimates of the bias parameters $B$ and $k$ is fairly straightforward, we saw in the STRIDE example that in practice, it is necessary to measure or estimate many more quantities to obtain estimates of the effective hazard ratio and the projected power.  All of these quantities come with their own uncertainties and could prove unstable when the interim data is examined, and systematically propagating uncertainty through the calculations is complicated by the lack of a closed-form expression for $H^{eff}$.

The variance inflation used for the effective sample sizes in the STRIDE example was based on model estimates.  There is a gap in methodology to estimate the design effect for complex survival data, but we are working on methods to be able to do so using an additive model.

Finally, we saw that STRIDE was a complex study and considerable effort was needed to apply these methods to it (such as accounting for clustering and bridging the gap between self-reported and adjudicated events).  Similarly, it may not be easy to generalize these methods to other studies.  However, we have set up a framework to consider, especially as the number of pragmatic trials increases and blinding of participants and interventionists is not feasible.

\section*{acknowledgements}
The STRIDE study was funded primarily by the Patient Centered Outcomes Research Institute (PCORI), with additional support from the National Institute on Aging (NIA), through Cooperative Agreement U01 AG048270.  Additional support was provided by CTSA Grant Number UL1 TR000142 from the National Center for Advancing Translational Science (NCATS) and by the Biostatistical Design and Analysis Core of the Boston Older Americans Independence Center (NIA Center Core Grant Number P30 AG031679).  NIA and NCATS are components of the National Institutes of Health (NIH).  The contents of this publication are solely the responsibility of the authors and do not necessarily represent the official view of NIH.

\textsf{SAS}, \textsf{SAS/STAT}, and all other \textsf{SAS} Institute Inc.\ product or service names are registered trademarks or trademarks of \textsf{SAS} Institute Inc.\ in the USA and other countries.

\section*{conflict of interest}
No authors reported a conflict of interest.

\section*{data availability statement}

The data that support the findings of this article are available from the corresponding author upon reasonable request.

\bibliography{refs.bib}
\renewcommand{\thepage}{S\arabic{page}}
\setcounter{page}{1}
\renewcommand{\figurename}{Supplementary~Figure}
\renewcommand{\thefigure}{S\arabic{figure}}
\setcounter{figure}{0}

\begin{figure}[p]
\centering
\includegraphics{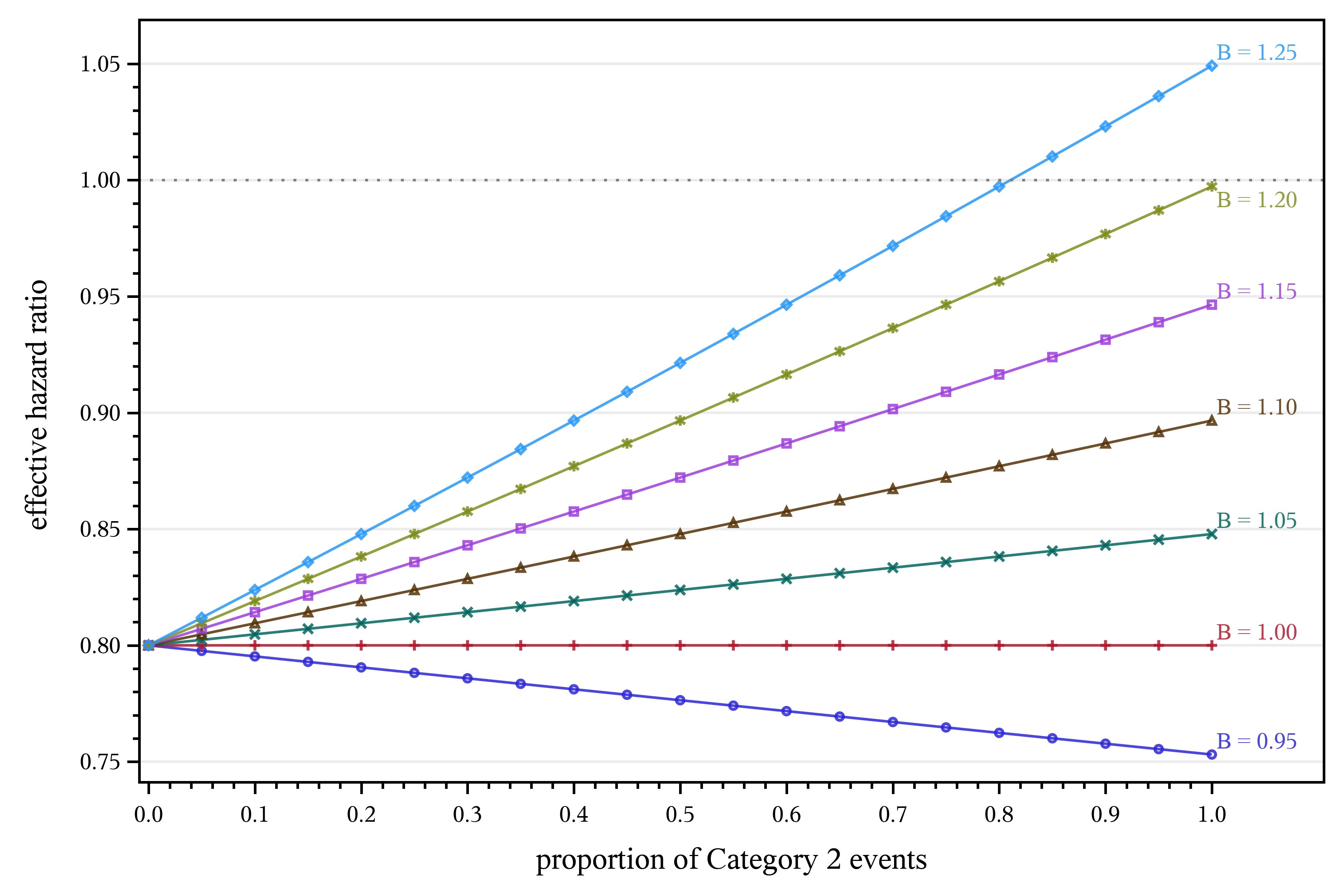}
\caption{Effects of inflation of events prone to bias ($B$; curve labels) and proportion of events prone to bias ($P$; x-axis) on the effective hazard ratio ($H^{eff}$). $N_C = N_I = 1611$, $T = 40 \text{ mo}$, $\mu = 0.5$, $H^{hyp} = 0.8$, $r^{prot} = 0.148$, $d = 0.025$.}
\label{fig:HvaryPB}
\end{figure}

\begin{figure}[p]
\centering
\includegraphics{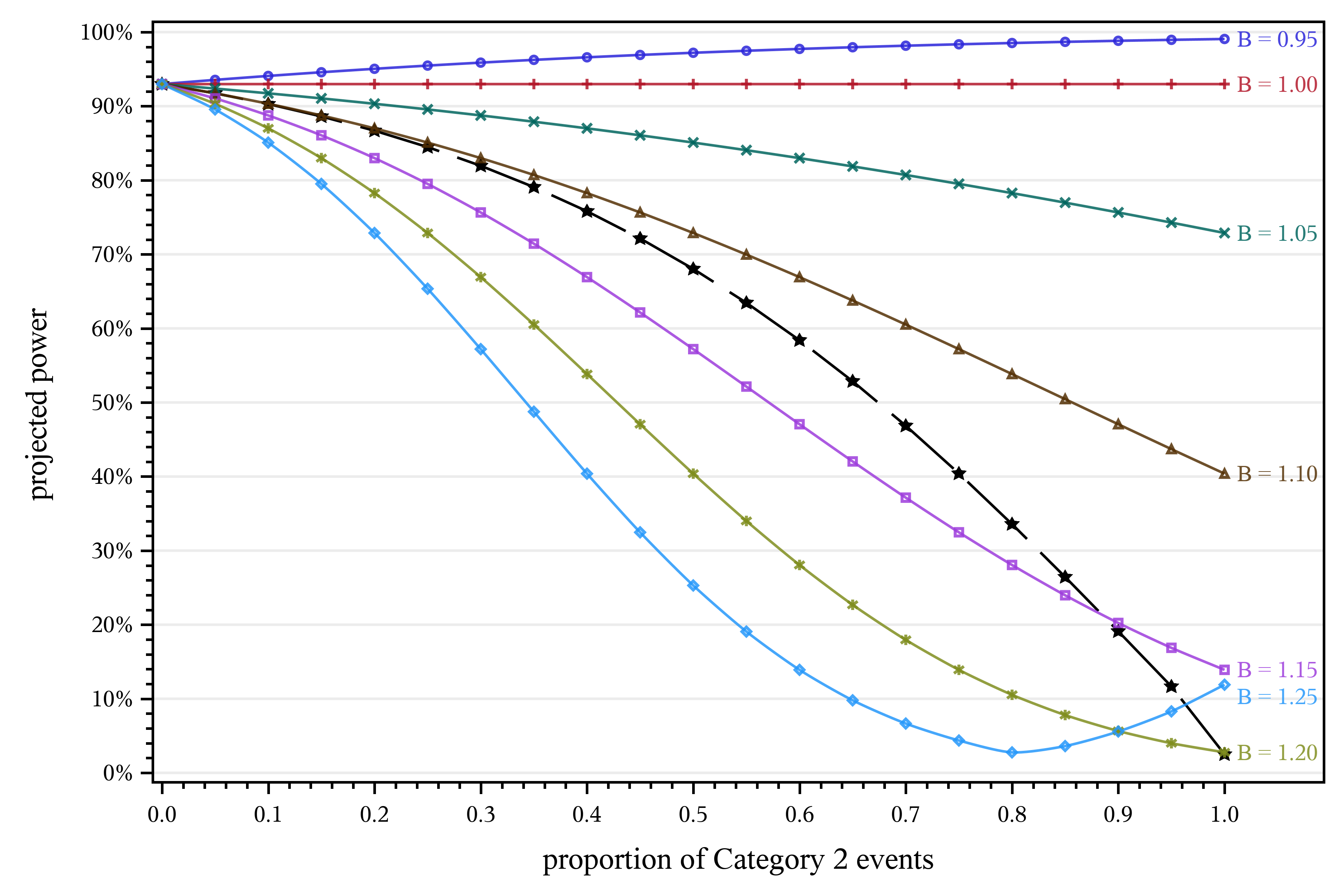}
\caption{Effects of inflation of events prone to bias ($B$; curve labels) and proportion of events prone to bias ($P$; x-axis) on projected power. The black dashed curve (no label) shows the projected power for the restricted definition; this is independent of $B$. $N_C = N_I = 1611$, $T = 40 \text{ mo}$, $\mu = 0.5$, $H^{hyp} = 0.8$, $r^{prot} = 0.148$, $d = 0.025$.}
\label{fig:PvaryPB}
\end{figure}

\begin{figure}[p]
\centering
\includegraphics{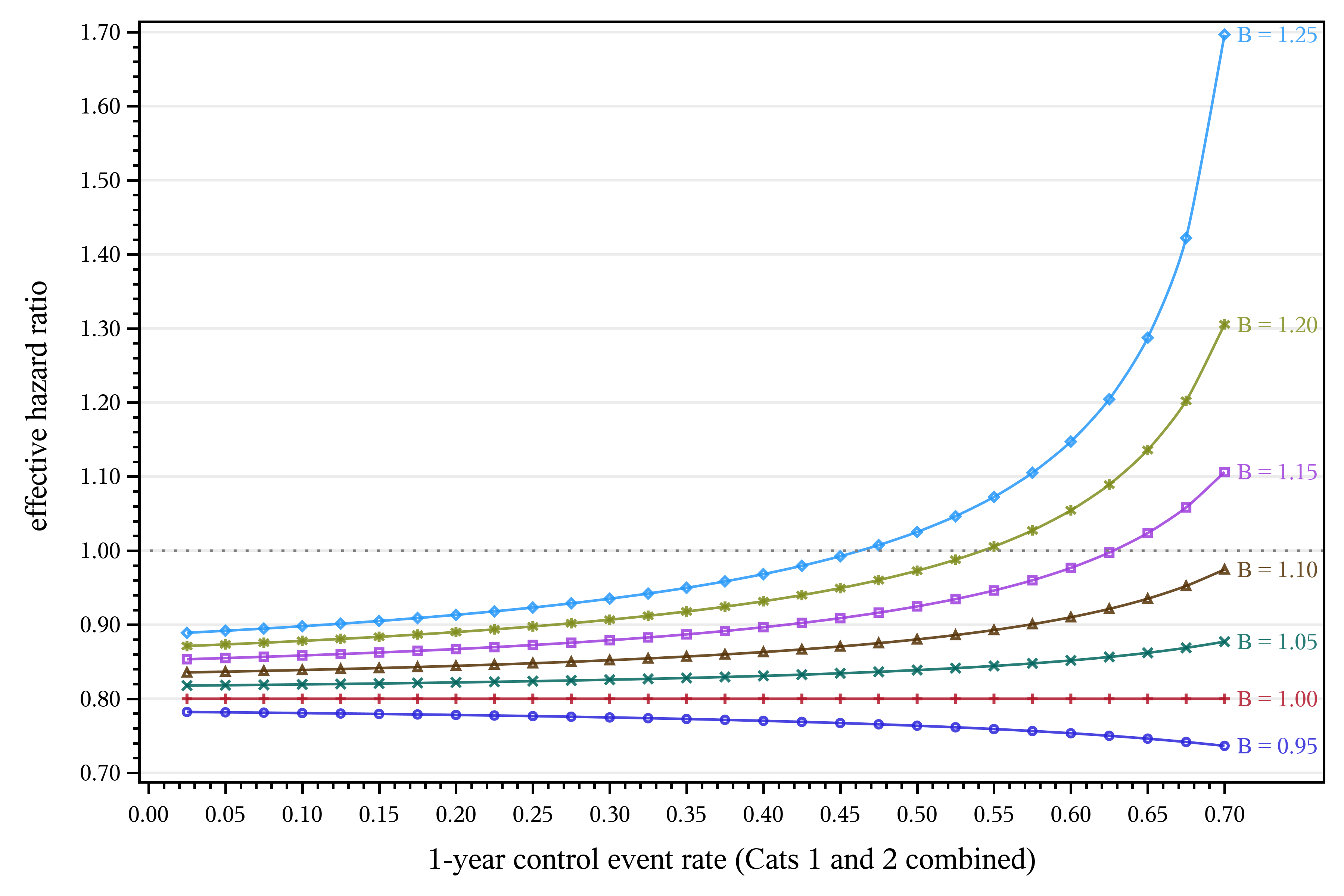}
\caption{Effects of inflation of events prone to bias ($B$; curve labels) and control event rate ($r^{prot}$; x-axis) on the effective hazard ratio ($H^{eff}$). $N_C = N_I = 1611$, $T = 40 \text{ mo}$, $\mu = 0.5$, $H^{hyp} = 0.8$, $d = 0.025$, $P = .432$.}
\label{fig:HvaryRB}
\end{figure}

\begin{figure}[p]
\centering
\includegraphics{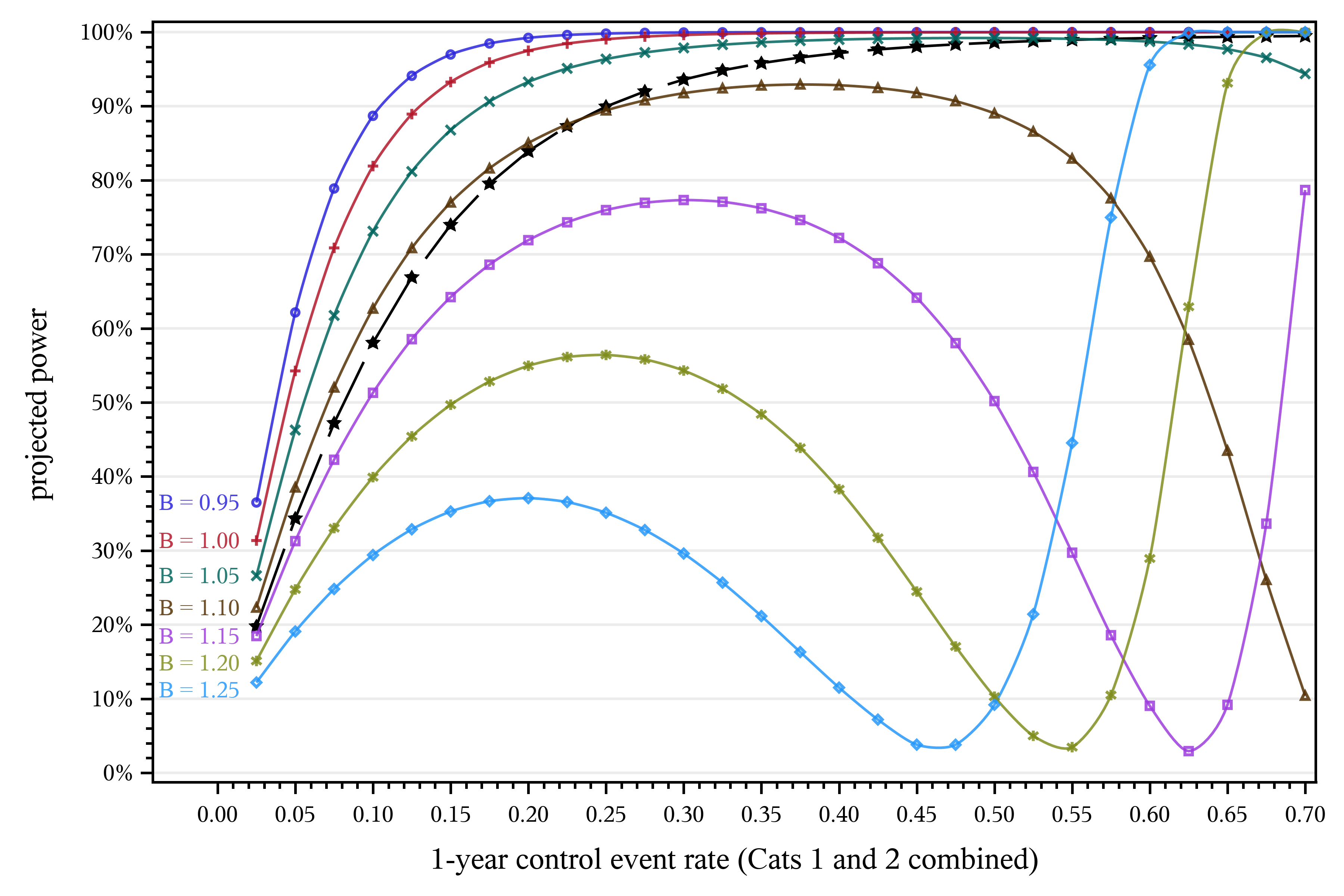}
\caption{Effects of inflation of events prone to bias ($B$; curve labels) and control event rate ($r^{prot}$; x-axis) on projected power. The black dashed curve (no label) shows the projected power for the restricted definition; this is independent of $B$. $N_C = N_I = 1611$, $T = 40 \text{ mo}$, $\mu = 0.5$, $H^{hyp} = 0.8$, $d = 0.025$, $P = .432$.}
\label{fig:PvaryRB}
\end{figure}

\begin{figure}[p]
\centering
\includegraphics{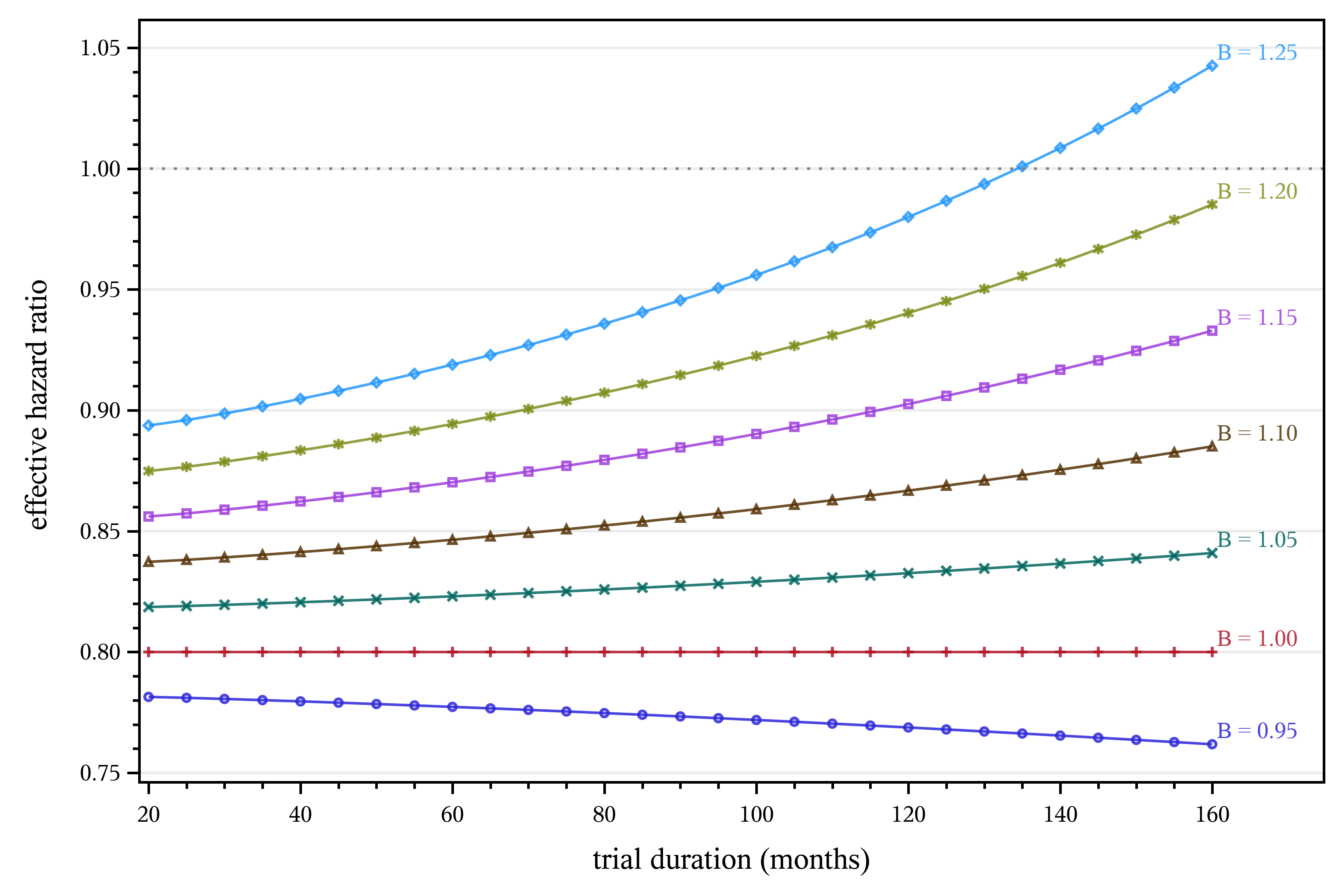}
\caption{Effects of inflation of  events prone to bias ($B$; curve labels) and total trial duration ($T$; x-axis) on the effective hazard ratio ($H^{eff}$). $N_C = N_I = 1611$, $\mu = \tfrac{20\text{ mo}}{T}$, $H^{hyp} = 0.8$, $r^{prot} = 0.148$, $d = 0.025$, $P = .432$.}
\label{fig:HvaryTB}
\end{figure}

\begin{figure}[p]
\centering
\includegraphics{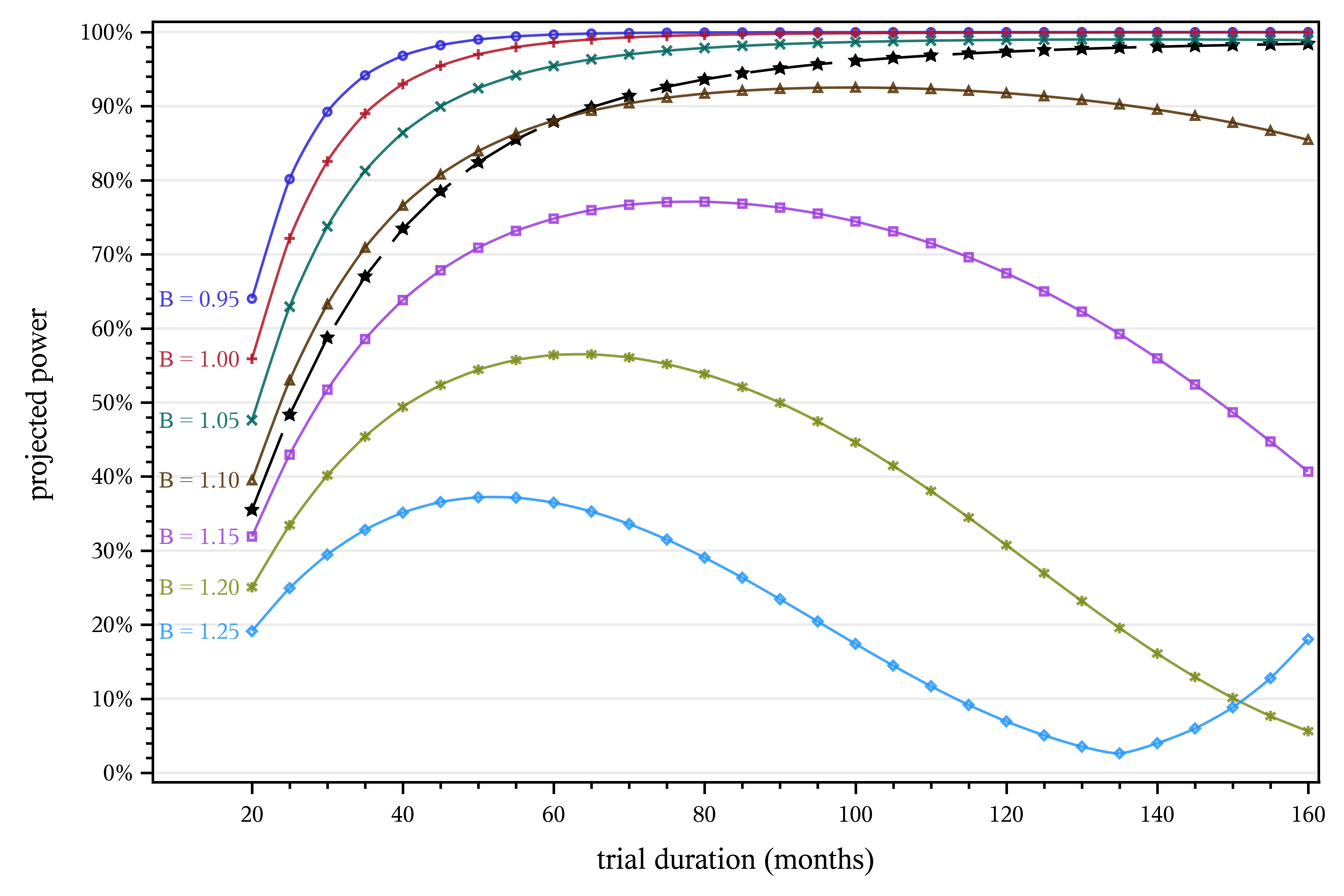}
\caption{Effects of inflation of events prone to bias ($B$; curve labels) and total trial duration ($T$; x-axis) on projected power. The black dashed curve (no label) shows the projected power for the restricted definition; this is independent of $B$. $N_C = N_I = 1611$, $\mu = \tfrac{20\text{ mo}}{T}$, $H^{hyp} = 0.8$, $r^{prot} = 0.148$, $d = 0.025$, $P = .432$.}
\label{fig:PvaryTB}
\end{figure}
\renewcommand{\thepage}{A\arabic{page}}
\setcounter{page}{1}
\renewcommand{\theequation}{A\arabic{equation}}
\setcounter{equation}{0}
\appendix

\section{Number of events} \label{app:numevents}

In this appendix, we derive the expected number of events for an outcome with hazard rate $\lambda$ in the presence of a competing risk with hazard rate $\gamma$.  Our starting point is the instantaneous hazard,
\begin{equation*}
f(t) = \lambda e^{-t(\lambda + \gamma)} \label{eq:lambda}.
\end{equation*}

Let the total trial duration be $T$, the enrollment period be a fraction $\mu$ of the overall trial, and  $\tau$ be the time a participant enters the trial; that participant's follow-up time is then $T-\tau$.  If we let $g(\tau)$ be the probability distribution of enrollment times (which can vary from 0 to $\mu T$), the probability of an individual having an outcome of interest is
\begin{align}
p &= \int_{0}^{\mu T} g(\tau) \left[ \int_{0}^{T-\tau} \lambda e^{-t(\lambda + \gamma)} \, dt \right] d\tau \nonumber \\
&= \int_{0}^{\mu T} g(\tau) \frac{\lambda}{\lambda + \gamma} \left[ -e^{-t(\lambda + \gamma)} \right]_{0}^{T-\tau} d\tau \nonumber \\
&= \frac{\lambda}{\lambda + \gamma} \int_{0}^{\mu T} g(\tau) \left[ 1 - e^{-(T-\tau)(\lambda + \gamma)} \right] d\tau \label{eq:nonuniform} .
\end{align}
The probability of an individual having an outcome of interest under uniform enrollment $g(\tau) = \frac{1}{\mu T}$ is then
\begin{align*}
p &= \frac{\lambda}{\lambda + \gamma} \int_{0}^{\mu T} \frac{1 - e^{-(T-\tau)(\lambda + \gamma)}}{\mu T} \, d\tau \nonumber \\
&= \frac{\lambda}{\lambda + \gamma} \left[ \frac{\tau}{\mu T} - \frac{e^{-(T-\tau)(\lambda + \gamma)}}{\mu T(\lambda + \gamma)} \right]_0^{\mu T} \nonumber \\
&= \frac{\lambda}{\lambda + \gamma} \left[ 1 - \frac{e^{-(1-\mu)T(\lambda + \gamma)}-e^{-T(\lambda + \gamma)}}{\mu T(\lambda + \gamma)} \right] .
\end{align*}

Assuming the distribution of events is binomial, the expected number of events is then 
\begin{equation*} 
E = Np = \frac{N\lambda}{\lambda + \gamma} Q(\lambda) \label{eq:AppE}
\end{equation*}
where $N$ is the total number of participants and 
\begin{equation} \label{eq:defQ}
Q(z) = 1 - \frac{e^{-(1-\mu)T(z + \gamma)}-e^{-T(z + \gamma)}}{\mu T(z + \gamma)}.
\end{equation}
Note that $N$ and $E$ can be written as $N = N_C + N_I$ and $E = E_C + E_I$ respectively, where $N_C$ and $E_C$ are the numbers in the control arm and $N_I$ and $E_I$ are the numbers in the intervention arm.

\section{Relationship between control and intervention events} \label{app:justifyconstantx}

Let the hazards for each event category in the control arm be $\lambda_1$, $\lambda_2$, and $\lambda_3$, and let their sum be $\xi = \lambda_1 + \lambda_2 +\lambda_3$.  Then, following the development of Appendix \ref{app:numevents} with $f_e(t) = \lambda_e e^{-t(\xi+\gamma)}$ for a specific event category $e$ ($e \in \left\{1,2,3\right\}$), the expected number of control events in Category $e$ is given by
\begin{equation*}
E_{Ce} = \frac{N_C\lambda_e}{\xi + \gamma} Q(\xi)
\end{equation*}
where $N_C$ is the number of individuals in the control arm and $Q(\xi)$ is defined in Equation \eqref{eq:defQ} . Note that $Q(\xi)$ does not depend on $e$.

Under the assumption of a constant hazard ratio  across event categories, for the intervention arm we have
\begin{gather*}
\forall e \in \lbrace 1,2,3 \rbrace: \lambda^I_e = H^{hyp} \lambda_e , \\
\xi^I = H^{hyp} \xi ;
\end{gather*}
then the expected number of type $e$ events in the intervention is given by
\begin{equation*}
E_{Ie} = \frac{N_I H^{hyp} \lambda_e}{H^{hyp}\xi + \gamma} Q(H^{hyp}\xi),
\end{equation*}
where $H^{hyp}$ is the hypothesized ratio and $N_I$ is the number of individuals in the intervention arm.

We can calculate the ratio of expected events in Category $e$ as
\begin{equation*}
\frac{E_{Ie}}{E_{Ce}} = \kappa = \frac{N_I}{N_C} \cdot H^{hyp} \cdot \frac{\xi + \gamma}{H^{hyp}\xi + \gamma} \cdot \frac{Q(H^{hyp}\xi)}{Q(\xi)},
\end{equation*}
which is independent of $e$.

\section{Estimation of $H^{eff}$} \label{app:solve4H}

We can rewrite Equation \eqref{EQ:HMONSTER} as
\begin{equation*}
\frac{\Lambda}{\Lambda+\gamma} \overline{Q}(x) = kZ,
\end{equation*}
where
\begin{equation} \label{eq:defQbar}
\overline{Q}(x) = 1 - \frac{e^{-(1-\mu)x}-e^{-x}}{\mu x} ,
\end{equation}

and 
\begin{align}
\Lambda &= H^{eff}\lambda , \nonumber \\
x &= T(\Lambda + \gamma), \nonumber \\
Z &= \frac{H^{hyp}\lambda}{H^{hyp}\lambda+\gamma} Q(H^{hyp}\lambda) .  \label{eq:Zdef}
\end{align}
Note that $\overline{Q}(x) = Q(H^{eff}\lambda)$; compare Equations \eqref{eq:defQ} and \eqref{eq:defQbar}.

We can then obtain $H^{eff}$ by finding the root $\Lambda^*$ (the expected intervention arm hazard after accounting for the bias) of the function
\begin{equation} \label{eq:fofBigL}
f(\Lambda,x(\Lambda)) = \frac{\Lambda}{\Lambda+\gamma} \overline{Q}(x) - kZ = 0
\end{equation}
and dividing the root by $\lambda$.

\subsection{Numerical approximation}

Equation \eqref{eq:fofBigL} is amenable to solution by Newton's method, whose iterations are $\Lambda_{n+1} = \Lambda_{n} - f(\Lambda_{n})/f'(\Lambda_{n})$.

We build up the total derivative of $f$ via the chain rule:
\begin{align*}
\frac{\partial f}{\partial \Lambda} &= \frac{\gamma}{(\Lambda+\gamma)^2}\overline{Q}(x) =  \frac{\gamma}{(\Lambda+\gamma)^2} \left[ 1 - \frac{e^{-(1-\mu)x}-e^{-x}}{\mu x} \right], \\
\frac{\partial f}{\partial x} &= \frac{\Lambda}{\Lambda+\gamma} \frac{d\overline{Q}}{dx} = \frac{\Lambda}{\Lambda+\gamma} \left( \frac{1}{\mu x^2} \right) \left\lbrace x\left[ (1-\mu)e^{-(1-\mu)x}-e^{-x} \right] + \left[ e^{-(1-\mu)x}-e^{-x} \right] \right\rbrace, \\
\frac{dx}{d\Lambda} &= T = \frac{x}{\Lambda+\gamma} ,
\end{align*}
and
\begin{align*}
\frac{df}{d\Lambda} &=  \frac{\partial f}{\partial \Lambda} + \frac{\partial f}{\partial x}\frac{dx}{d\Lambda} \nonumber \\
&= \frac{1}{\mu x (\Lambda + \gamma)^2} \left\lbrace \gamma \left[ \mu x - e^{-(1-\mu)x} + e^{-x} \right] + \Lambda \left[ x\left( (1-\mu)e^{-(1-\mu)x}-e^{-x} \right) + e^{-(1-\mu)x} - e^{-x} \right] \right\rbrace .
\end{align*}

The second-order approximation given in Equation \eqref{eq:lambda2} below is frequently a suitable starting point $\Lambda_0$.

\subsection{Analytical approximation}

\subsubsection{Taylor series expansion of $\overline{Q}$} \label{app:taylor}

First, we note that if we take the binomial expansion
\begin{equation*}
(a + b)^n = \sum_{k=0}^n \binom{n}{n-k} a^{n-k} b^k = \sum_{k=0}^n \frac{n!}{k!(n-k)!} a^{n-k} b^k
\end{equation*}
and make the substitutions $a = 1$ and $b = -\mu$, some algebra leads to
\begin{align}
\frac{1 - (1-\mu)^{n+1}}{\mu(n+1)!} &= \frac{1}{\mu(n+1)!} \left[ 1 - \sum_{k=0}^{n+1} \frac{(n+1)!}{k!(n+1-k)!} (-\mu)^k \right] \nonumber \\
&= \frac{1}{\mu(n+1)!} \left[ - \sum_{k=1}^{n+1} \frac{(n+1)!}{k!(n+1-k)!} (-\mu)^k \right] \nonumber \\
&= \sum_{k=1}^{n+1} \frac{(-\mu)^{k-1}}{k!(n-k+1)!} \nonumber \\
&= \sum_{k=0}^n \frac{(-\mu)^k}{(k+1)!(n-k)!} .  \label{eq:binompiece}
\end{align}
We also recall from pre-calculus that if two infinite series converge, their sum can be expressed as another infinite series:
\begin{equation} \label{eq:remedial}
\sum_{n=0}^{\infty} a_n + \sum_{n=0}^{\infty} b_n = \sum_{n=0}^{\infty} \left( a_n + b_n \right) .
\end{equation}

We now build up the Taylor series expansion of $\overline{Q}(x)$ from the well-known expansion of $e^x$,
\begin{equation*}
e^x = \sum_{n=0}^{\infty} \frac{x^n}{n!} ,
\end{equation*}
which converges for all finite $x$.
\begin{align*}
e^{-(1-\mu)x} - e^{-x} &= \sum_{n=0}^{\infty} \frac{[-(1-\mu)x]^n}{n!} - \sum_{n=0}^{\infty} \frac{(-x)^n}{n!} \\
&= \sum_{n=0}^{\infty} \frac{1}{n!} \left\{ [ -\left( 1-\mu \right) x ]^n - (-x)^n \right\} \qquad \text{by \eqref{eq:remedial}} \nonumber \\
&= \sum_{n=0}^{\infty} \frac{(-1)^n \left[ (1-\mu)^n - 1 \right]}{n!} x^n \nonumber \\
&= 0 + \sum_{n=1}^{\infty} \frac{(-1)^n \left[ (1-\mu)^n - 1 \right]}{n!} x^n . \nonumber
\end{align*}
\begin{align*}
\frac{e^{-(1-\mu)x}-e^{-x}}{\mu x} &= \sum_{n=1}^{\infty} \frac{(-1)^n \left[ (1-\mu)^n - 1 \right]}{\mu n!} x^{n-1} \nonumber \\
&= \sum_{n=1}^{\infty} \frac{(-1)^{n-1} \left[ 1 - (1-\mu)^n \right]}{\mu n!} x^{n-1} \nonumber \\
&= \sum_{n=0}^{\infty} \frac{(-1)^{n} \left[ 1 - (1-\mu)^{n+1} \right]}{\mu(n+1)!} x^n \nonumber \\
&= 1 + \sum_{n=1}^{\infty} \frac{(-1)^{n} \left[ 1 - (1-\mu)^{n+1} \right]}{\mu(n+1)!} x^n . \nonumber
\end{align*}
Finally,
\begin{alignat*}{2}
\overline{Q}(x) = 1 - \frac{e^{-(1-\mu)x}-e^{-x}}{\mu x} &= - \sum_{n=1}^{\infty} \frac{(-1)^{n} \left[ 1 - (1-\mu)^{n+1} \right]}{\mu(n+1)!} x^n &\quad& \nonumber \\
&= \sum_{n=1}^{\infty} \frac{(-1)^{n+1} \left[ 1 - (1-\mu)^{n+1} \right]}{\mu(n+1)!} x^n && \nonumber \\
&= \sum_{n=1}^{\infty} \left[ \sum_{k=0}^n \frac{(-1)^{n+k+1} \mu^k}{(k+1)!(n-k)!} \right] x^n && \text{by~\eqref{eq:binompiece}}   \\
&= \left( 1-\tfrac{\mu}{2} \right) x - \left( \tfrac{1}{2} - \tfrac{\mu}{2} + \tfrac{\mu^2}{6} \right) x^2 + \mathcal{O}(x^3) . &&
\end{alignat*}

\subsubsection{First-order approximation}

To first order,
\begin{equation*}
\overline{Q}_{[1]}(x) = \left( 1-\tfrac{\mu}{2} \right) x = \left( 1-\tfrac{\mu}{2} \right) T(\Lambda + \gamma) ,
\end{equation*}
and Equation \eqref{eq:fofBigL} becomes
\begin{equation*}
\frac{\Lambda}{\Lambda+\gamma} \left( 1-\tfrac{\mu}{2} \right) T(\Lambda + \gamma) - kZ = H^{eff}\lambda T \left( 1-\tfrac{\mu}{2} \right) - kZ = 0
\end{equation*}
with solution
\begin{align*}
H^{eff}_{[1]} &= \frac{kZ}{\lambda T \left( 1-\tfrac{\mu}{2} \right)} \nonumber \\
&= \frac{k}{\lambda T \left( 1-\tfrac{\mu}{2} \right)} \frac{H^{hyp}\lambda}{H^{hyp}\lambda+\gamma} Q(H^{hyp}\lambda) \nonumber \\
&= \frac{kH^{hyp}}{\left( 1-\tfrac{\mu}{2} \right) T (H^{hyp}\lambda+\gamma)} \left[ 1 - \frac{e^{-(1-\mu)T(H^{hyp}\lambda + \gamma)}-e^{-T(H^{hyp}\lambda + \gamma)}}{\mu T(H^{hyp}\lambda + \gamma)} \right] .
\end{align*}

If we also look at a first-order approximation for $Q(H^{hyp}\lambda)$, we find that to lowest nontrivial order,
\begin{equation*}
H^{eff}_{[1]} \approx \frac{kH^{hyp}}{\left( 1-\tfrac{\mu}{2} \right) T (H^{hyp}\lambda+\gamma)} \left( 1-\tfrac{\mu}{2} \right) T(H^{hyp}\lambda + \gamma) = kH^{hyp}
\end{equation*}
as one might naively expect.

\subsubsection{Second-order approximation}

To second order,
\begin{align*}
\overline{Q}_{[2]}(x) &= \left( 1-\tfrac{\mu}{2} \right) x - \left( \tfrac{1}{2} - \tfrac{\mu}{2} + \tfrac{\mu^2}{6} \right) x^2 \nonumber \\
&= \left( 1-\tfrac{\mu}{2} \right) T(\Lambda + \gamma) - \left( \tfrac{1}{2} - \tfrac{\mu}{2} + \tfrac{\mu^2}{6} \right) T^2 (\Lambda + \gamma)^2,
\end{align*}
and Equation \eqref{eq:fofBigL} becomes
\begin{equation*}
\frac{\Lambda}{\Lambda+\gamma} \left[ \left( 1-\tfrac{\mu}{2} \right) T(\Lambda + \gamma) - \left( \tfrac{1}{2} - \tfrac{\mu}{2} + \tfrac{\mu^2}{6} \right) T^2 (\Lambda + \gamma)^2 \right] - kZ = 0 ,
\end{equation*}
which can be rearranged into the standard quadratic form
\begin{equation*}
\Lambda^2 - \left[ \frac{1-\tfrac{\mu}{2}}{T\left( \tfrac{1}{2} - \tfrac{\mu}{2} + \tfrac{\mu^2}{6} \right)} - \gamma \right] \Lambda + \frac{kZ}{T^2 \left( \tfrac{1}{2} - \tfrac{\mu}{2} + \tfrac{\mu^2}{6} \right)} = 0
\end{equation*}
(since $T$ is assumed to be positive and $\tfrac{1}{2} - \tfrac{\mu}{2} + \tfrac{\mu^2}{6}$ has no real zeroes). If we introduce the quantities
\begin{equation*} \label{eq:thetadef}
\theta = \frac{1}{2} \left[ \frac{1-\tfrac{\mu}{2}}{T\left( \tfrac{1}{2} - \tfrac{\mu}{2} + \tfrac{\mu^2}{6} \right)} - \gamma \right] = \frac{1}{2} \left[ \frac{3}{T} \left(\frac{2-\mu}{\mu^2 - 3\mu +3}\right) - \gamma \right]
\end{equation*}
and
\begin{equation*} \label{eq:phidef}
\phi =  \frac{1}{T} \sqrt\frac{kZ}{\tfrac{1}{2} - \tfrac{\mu}{2} + \tfrac{\mu^2}{6}} = \frac{1}{T} \sqrt\frac{6kZ}{\mu^2 - 3\mu +3} ,
\end{equation*}
this reduces to
\begin{equation*}
\Lambda^2 - 2\theta\Lambda + \phi^2 = 0
\end{equation*}
with solutions
\begin{equation*}
\Lambda = \tfrac{1}{2} \left( 2\theta \pm \sqrt{4\theta^2 - 4\phi^2} \right) = \theta \left( 1 \pm \sqrt{1 - \phi^2/\theta^2} \right) .
\end{equation*}
To decide which root to use, we require that as more intervention events are observed, the intervention event rate $\Lambda$ increases (rather than decreases).  Substituting Equation \eqref{eq:Zdef} into Equation \eqref{eq:defHeff}, we see that $E_I^{obs} = N_I^{*}kZ$, which is proportional to $\phi^2$, so we want the root where $\Lambda$ increases with $\phi^2$, i.e., where $\frac{\partial \Lambda}{\partial (\phi^2)} > 0$.  Since
\begin{equation*}
    \frac{\partial}{\partial (\phi^2)} \left[ \theta \left( 1 \pm \sqrt{1 - \phi^2/\theta^2} \right) \right] = \pm \frac{-1}{2\theta \sqrt{1 - \phi^2/\theta^2}},
\end{equation*}
we need to select the minus sign rather than the plus sign to satisfy this condition, and
\begin{equation} \label{eq:lambda2}
\Lambda = \theta \left( 1 - \sqrt{1 - \phi^2/\theta^2} \right) .
\end{equation}
Finally, we divide through by $\lambda$ to obtain the second-order approximation to the effective hazard,
\begin{equation} \label{eq:approx2}
H^{eff}_{[2]} = \frac{\Lambda}{\lambda} = \frac{\theta}{\lambda} \left( 1 - \sqrt{1 - \phi^2/\theta^2} \right) .
\end{equation}

Since this approximation is based on a Taylor series expansion in powers of $x = T(\Lambda + \gamma)$, we would expect it to break down as $\Lambda T$ or $\gamma T$ grows large.  And we see that if $\theta$ becomes negative, the effective hazard ratio also becomes negative, which is impossible, so the approximation is clearly invalid unless $\theta \geq 0$, or
\begin{equation*}
\gamma T \leq \frac{3(2 - \mu)}{\mu^2 - 3\mu + 3} .
\end{equation*}
Further, Equation \eqref{eq:approx2} has real solutions only when $\theta^2 \geq \phi^2$; since we know $\theta$ must be nonnegative and $\phi$ is defined to be positive, we can further tighten our sanity constraint to $\theta \geq \phi$, or
\begin{equation*}
\gamma T \leq \frac{3(2 - \mu)}{\mu^2 - 3\mu + 3} - \sqrt{\frac{24kZ}{\mu^2 - 3\mu + 3}} .
\end{equation*}
In simulations where this constraint is violated, $\Lambda_0 = \theta$ has worked as a starting point for numerical estimation.

\section{Computing hazards from cause-specific rates} \label{app:realrates}

With a control outcome hazard rate $\lambda$ and a (competing) death hazard rate $\gamma$, the instantaneous event and death hazards are
\begin{align*}
f_e(t) &= \lambda e^{-t(\lambda + \gamma)} , \\
f_d(t) &= \gamma e^{-t(\lambda + \gamma)} .
\end{align*}

Let $r$ and $d$ be our observed cause-specific cumulative control event and death rates at 12 months.  Then
\begin{align}
r &= \int_{0}^{12\text{ mo}} \lambda e^{-t(\lambda + \gamma)} dt = \frac{\lambda}{\lambda + \gamma} \left[ 1 - e^{-(12\text{ mo})(\lambda + \gamma)} \right] , \label{eq:r12mo} \\
d &= \int_{0}^{12\text{ mo}} \gamma e^{-t(\lambda + \gamma)} dt = \frac{\gamma}{\lambda + \gamma} \left[ 1 - e^{-(12\text{ mo)}(\lambda + \gamma)} \right] . \nonumber
\end{align}
Solving each of these equations for $\left[1 - e^{-(12\text{ mo})(\lambda + \gamma)} \right] / \left(\lambda + \gamma\right)$ and equating the results, we find
\begin{equation*}
\frac{1 - e^{-(12\text{ mo})(\lambda + \gamma)}}{\lambda + \gamma} = \frac{r}{\lambda} = \frac{d}{\gamma} ,
\end{equation*}
giving us a relation between $\lambda$ and $\gamma$,
\begin{equation} \label{eq:lambda2gamma}
\gamma = \lambda \left( \frac{d}{r} \right) ,
\end{equation}
and the substitutions
\begin{equation*}
\lambda + \gamma = \lambda \left( \frac{r+d}{r} \right)
\end{equation*}
and
\begin{equation*}
\frac{\lambda}{\lambda + \gamma} = \frac{r}{r+d} .
\end{equation*}

Making these substitutions in Equation \eqref{eq:r12mo}, we obtain
\begin{align*}
r &= \frac{r}{r+d} \left[ 1 - e^{-(12\text{ mo})\lambda (r+d)/r} \right] \\
e^{-\lambda(12\text{ mo})(r+d)/r} &= 1-(r+d) \\
-\lambda(12\text{ mo})\left( \frac{r+d}{r} \right) &= \ln (1-r-d)
\end{align*}
and finally
\begin{equation} \label{eq:h1}
\lambda = - \left( \frac{r}{r+d} \right) \frac{\ln(1-r-d)}{12\text{ mo}} .
\end{equation}

From Equation \eqref{eq:lambda2gamma}, we then also have
\begin{equation} \label{eq:h2}
\gamma = - \left( \frac{d}{r+d} \right) \frac{\ln(1-r-d)}{12\text{ mo}} .
\end{equation}

\section{Confidence intervals for bias parameters} \label{app:calcCI}

Recall that for a function $f(x_1,x_2,\dots,x_n)$ where the variances and covariances of the arguments are known, the function's variance (to first order in the argument variances) is
\begin{equation}
\sigma^2_f = \sum_i \left( \frac{\partial f}{\partial x_i} \right)^2 \sigma^2_{x_i} + \sum_{i \neq j} \frac{\partial f}{\partial x_i} \frac{\partial f}{\partial x_j} \sigma_{x_i x_j} .
\label{eq:propagate}
\end{equation}
We rely on this to estimate the variances of functions of more than one variable.

Our bias estimates flow from the three proportions $P$, $\rho_I$, and $\rho_C$, whose variances are all of the form $p(1-p)/D$, where $D$ is the denominator of the proportion:
\begin{align*}
\sigma^2_{\rho_I} &= \frac{E^{obs}_{I2}E^{obs}_{I3}}{\left( E^{obs}_{I2} + E^{obs}_{I3} \right)^3} , \\
\sigma^2_{\rho_C} &= \frac{E_{C2}E_{C3}}{\left( E_{C2} + E_{C3} \right)^3} , \\
\sigma^2_P &= \frac{E_{C1}E_{C2}}{\left( E_{C1} + E_{C2} \right)^3} .
\end{align*}
We assume no covariance among them.

$B$ is the ratio of $\rho_I$ to $\rho_C$; Equation (\ref{eq:bdef}), Equation (\ref{eq:propagate}), and some algebra yield
\begin{equation*}
\sigma^2_B = B^2 \left( \frac{\sigma^2_{\rho_I}}{\rho^2_I} + \frac{\sigma^2_{\rho_C}}{\rho^2_C} \right) .
\end{equation*}

$k$ in turn depends on $B$ and $P$; Equations (\ref{eq:kdef}) and (\ref{eq:propagate}) plus more algebra yield
\begin{equation*}
    \sigma^2_k = (B-1)^2 \sigma^2_P + P^2 \sigma^2_B .
\end{equation*}

All the bias information needed for further calculation is encapsulated in $k$, and the number of observed intervention events $E_I^{obs}$, the effective hazard ratio $H^{eff}$, and the estimated power under ascertainment bias $1-\beta^{prot}$ are monotonic implicit functions of $k$.  If the trial parameters are treated as given, the procedure used to determine these quantities from $k$ can also be used to determine the bounds of their confidence intervals from the bounds of the confidence interval around $k$.  (For example, if we let $\mathcal{F}$ be the implied function relating $E_I^{obs}$ to $k$, i.e., $E_I^{obs} = \mathcal{F}(k)$, then the bounds of the confidence interval around $E_I^{obs}$ are $\mathcal{F}(k^{lower})$ and $\mathcal{F}(k^{upper})$; the same applies to $H^{eff}$ and $1-\beta^{prot}$.)

In practice, however, the hazards and effective sample sizes are based on observed outcome, death, and withdrawal rates and estimated design effects and so are also uncertain.  Since there is no closed-form expression for $H^{eff}$, confidence intervals for the effective hazard ratio and projected power can only be obtained via simulation.

\end{document}